%% file: Main.tex
\documentclass[lineno]{jfm}
\usepackage{graphicx}
\usepackage{epstopdf,epsfig}
\usepackage{newtxtext}
\usepackage{newtxmath}
\usepackage{natbib}
\usepackage{hyperref}
\usepackage{xcolor}
\usepackage{amsmath}
\usepackage{amsfonts}
\usepackage{pgfplots}
\usepackage{breqn}
\pgfplotsset{compat=1.18}
\usepackage{algpseudocode}
\usepackage{algorithm}
\usepackage{varwidth}
\usepackage{float}
\usepackage{academicons}
\usepackage{tikz}
\usepackage{siunitx}
\usepackage{float}
\usepackage{graphicx} 
\usepackage{multirow} 
\usepackage{comment} 
\usepackage{pgfplots} 
\usepackage{adjustbox}  
\usepackage{xcolor}    
\usepackage{ragged2e} 
\usepackage{multicol} 
\usepackage{rotating}
\usepackage{caption}
\usepackage{subcaption}
\usepackage{booktabs}
\usepackage{nth}
\usepackage{textpos}
\usepackage{braket} 
\usepackage{orcidlink}
\usetikzlibrary{arrows}
\usetikzlibrary{arrows.meta}
\usetikzlibrary{fit,positioning}
\usetikzlibrary{shapes.geometric}
\providecommand{\norm}[1]{\lVert#1\rVert}
\newcommand{\abs}[1]{\left|{#1}\right|}
\newcommand\Algphase[1]{%
\vspace*{-.7\baselineskip}\Statex\hspace*{\dimexpr-\algorithmicindent-2pt\relax}\rule{\textwidth}{0.4pt}%
\Statex\hspace*{-\algorithmicindent}\textbf{#1}%
\vspace*{-.7\baselineskip}\Statex\hspace*{\dimexpr-\algorithmicindent-2pt\relax}\rule{\textwidth}{0.4pt}%
\algrenewcommand\algorithmicrequire{\textbf{Input:}}
\algrenewcommand\algorithmicensure{\textbf{Output:}}
}

\definecolor{orcidlogocol}{HTML}{A6CE39}
\hypersetup{
    colorlinks = true,
    urlcolor   = blue,
    citecolor  = black,
}

\newcommand{\RomanNumeralCaps}[1]
\linenumbers
\title{Self-tuning model predictive control for wake flows}
\author{Luigi Marra\aff{1}
  \corresp{\email{lmarra@pa.uc3m.es}},
  Andrea Meilán-Vila\aff{2},
 \and Stefano Discetti\aff{1}}
\affiliation{\aff{1}Department of Aerospace Engineering, Universidad Carlos III de Madrid, Av. de la Universidad 30, Leganés, 28911, Madrid, Spain
\aff{2}Department of Statistics, Universidad Carlos III de Madrid, Av. de la Universidad 30, Leganés, 28911, Madrid, Spain}
\shortauthor{L. Marra, A. Meilán-Vila and S. Discetti}
\begin{document}
\maketitle
%
%
%
\begin{abstract}\nolinenumbers
\label{sec:Abstract}This study presents a noise-robust closed-loop control strategy for wake flows employing model predictive control. The proposed control framework involves the autonomous offline selection of hyperparameters, eliminating the need for user interaction. To this purpose, Bayesian optimization maximises the control performance, adapting to external disturbances, plant model inaccuracies and actuation constraints. The noise robustness of the control is achieved through sensor data smoothing based on local polynomial regression. The plant model can be identified through either theoretical formulation or using existing data-driven techniques. In this work we leverage the latter approach, which requires minimal user intervention. The self-tuned control strategy is applied to the control of the wake of the fluidic pinball, with the plant model based solely on aerodynamic force measurements. The closed-loop actuation results in two distinct control mechanisms: boat tailing for drag reduction and stagnation point control for lift stabilization. The control strategy proves to be highly effective even in realistic noise scenarios, despite relying on a plant model based on a reduced number of sensors.
\end{abstract}
%
%
%
%
%
%
\nolinenumbers
\section{Introduction}
 \label{sec:Introduction}

Prediction and control of fluid flows to pursue a specific objective is a highly compelling research area \citep{GadElHak2000FlowControl}. Flow control offers wide-ranging practical applications in diverse fields, including vehicle dynamics, aircraft and marine transportation, meteorology, energy production from water and wind, combustion and chemical processes, and more \citep{duriez2017machine}. The goals of fluid flow control generally encompass, among others, drag reduction, control of separation and transition, lift or mixing enhancement. In recent years, drag reduction has received considerable attention due to its notable impact on the environmental footprint of transportation means \citep{Green2003AviationEnvirChallenge}.

In the last decades, active flow control has garnered increasing attention. This technique can be implemented in open-loop and closed-loop configurations. The former involves predetermining the actuation law irrespective of the system state, thus simplifying its application. Notable examples include wake control of bluff bodies \citep{Blackburn1999OscCyl,Cetiner2001StreamwiseOscCyl,Thiria2006WakeCyl,Parkin2014OpenLoopBluffBody}, open-cavity flows \citep{Sipp2012OpenLoopCavity,Little2007OpenLoopCavity,Nagarajan2018OpenLoopCavity} and heat transfer \citep{castellanos2022reducing} among others. However, open-loop control's effectiveness is limited in unstable flow stabilization, responding to changes in the controlled system parameters or dealing with external disturbances. On the contrary, closed-loop implementations (also referred to as reactive control) involve feeding the control law by the knowledge of the state.
This approach offers greater flexibility and adaptability \citep{Brunton2015ClosedLoopControl}. Experimental evidence demonstrates the superior performance of closed-loop over open-loop control; see e.g. \citet{Pinier2007ClosedLoopwing} or \citet{Shimomura2020CLwing}.

The identification of control laws requires adequate knowledge of the system dynamics and its response to control inputs. In fluid dynamics, model-based techniques have traditionally been utilised to obtain this information, proving successful in various scenarios \citep{Kim2007LinearFlowControl}.  Examples of applications include transition delay in spatially evolving wall-bounded flows \citep{Monokrousos2008TransitionDelay, Chevalier2007DelayTransition, Tol2019ModelTransitionDelay}, cavity flow control \citep{Rowley2006ControlOpenCavity,Illingworth2011ControlFlowReson}, separation control on a low-Reynolds-number airfoil \citep{Ahuja2007ControlLEvortices}, wake stabilization of cylinders \citep{Schumm1994ControlWakeBluffBody,Gerhard2003ModelBasedGalerkVortexShed,Tadmor2011WakeControl}, skin-friction drag reduction \citep{Cortelezzi1998ControlTransition,Lee2001ModelBasedTurbulentDragReduction,Kim2011ControlWallTurbulenceDragReduction}. However, the identification of efficient analytical control laws faces an important challenge in the presence of complex nonlinear multiscale dynamics.

In recent years, model-free techniques have gained popularity, driven by advancements in hardware and the increasing efficiency of data-driven and machine-learning algorithms. Examples of model-free techniques include genetic algorithms in jet mixing optimization \citep{Koumoutsakos2001GAJetMixing, Wu2018MLJet}, wake flows \citep{Poncet2005ESwakes, Raibaudo2020MLpinball}, separation control \citep{Gautier2015closed} and combustion noise \citep{Buche2002MOEAcombustion}. Reinforcement learning (RL) has also recently gained popularity, with successful applications in the control of bluff body wakes \citep{rabault2019ANNandRLflowcontrol, Fan2020reinforcement, castellanos2022machine} and natural convection \citep{Beintema2020RLRBconvection}. Despite the encouraging results of such model-free techniques, their effectiveness is limited by the need for large datasets. 

Within model-based techniques, model predictive control (MPC) offers interesting features to deal with the challenges of fluid flow control. Model predictive control is based on the idea of receding horizon control. It has found application in the industry since the 1980s \citep{Qin2003SurveyMPC}, in particular with extensive use in refineries and the petrochemical industry \citep{Lee2011MPCreview}.
Model predictive control has demonstrated excellent performance in controlling complex systems with constraints, strong nonlinearities, and time delays \citep{Henson1998NMCPstatus,Allgower2004NMPCtheoryandapp,Camacho2013MPC,Grune2017NMPC}. Therefore, it is particularly appropriate for complex systems that challenge traditional linear controllers \citep{Corona2008CruiseContrMPC}. The method requires the identification of a model of the system dynamics capable of predicting its behaviour under exogenous inputs. The optimal control is determined through the iterative solution of an optimization problem within a prediction window, aiming to minimise a user-defined cost function that considers the distance of the system state from the control target. 
Moreover, model predictive control allows for the straightforward implementation of hard constraints, such as hardware limitations, distinguishing it from classical control approaches. Model predictive control has been successfully applied in the control of complex fluid systems, see, e.g. \citet{Collis2000LESandTurbControl,Bewley2001DNSpredictiveControlTurbulence,Bieker2019DeepMPC, Sasaki2020MPCcyl,Morton2018CylMPC, Peitz2020DataDrivenKoopmanMPC}.  
A crucial aspect of MPC implementation is achieving a proper balance among the terms of the loss function. The user needs to select weights (referred to as hyperparameters) for the loss, considering factors like closeness to the target, cost of the action and other application-tailored constraints. This choice has a clear impact on the final performance. In flow control applications this process traditionally relies on user experience, which poses the risk of suboptimal choices.

Bayesian optimization (BO) or RL techniques have demonstrated excellent results in hyperparameter tuning, particularly in the fields of autonomous driving and robotics \citep{Edwards2021LearningBasedMPC,Frohlich2022LearningBasedMPC,Bohn2023LearningBasedMPC}. A comprehensive review in this area can be found in \citet{Hewing2020LearningMPC}. However, in the application of nonlinear MPC to flow control, examples are scarce, and the choice of MPC parameters is often guided by trial error and intuition. This approach risks falling into suboptimal configurations that may not adequately account for the different degrees of fidelity in the terms involved in the loss function. This issue is particularly relevant in fluid mechanics, where the uncertainty in predicting the plant behaviour and the measured state/control actions should play a role in the parameter selection process. Unfortunately, an analytical formulation is elusive in most cases.

Moreover, as a closed-loop strategy, the implementation of MPC requires feedback, consisting of time-sampled measurements of a feature of the system to be controlled. In real control scenarios, this sampling is often affected by measurement noise, which can compromise control decision making. Thus, suitable smoothing techniques are necessary to enhance noise robustness. In time series analysis a non-parametric statistical technique called local polynomial regression (LPR) proves particularly effective in this task. Local polynomial regression estimates the regression function of sensor outputs and their time derivatives without assuming any prior information. Applications of LPR for control purposes are described in works such as \citet{Steffen2010LPRclosedloopRobotics} or \citet{Ouyang2018lprcontr}.

In this paper, we propose a fully automatic architecture that self-tunes control and optimization process parameters with minimal user input. Our MPC framework adapts to different levels of noise and/or limited state knowledge. The methodology builds upon offline black-box optimization via Bayesian methods for hyperparameter tuning. Furthermore, we discuss the robustness enhancement to noise using an online LPR. The effectiveness of the control algorithm is evaluated through its application to the control of the wake of the fluidic pinball \citep{Deng2020ROMpinball} in the chaotic regime. Although not strictly needed, plant identification is also data driven. In this work, nonlinear system identification is performed using the sparse identification of nonlinear dynamics with control \citep[SINDYc, ][]{Brunton2016SINDYc}.

The paper is organised as follows. Section \ref{sec:Methodology} provides a description of the methodology, emphasizing the mathematical tools and the MPC framework employed. Additionally, this section includes specific details regarding the chosen test case for illustration purposes. The results of the control application, along with their interpretation are provided in \S~\ref{sec:Results}. Finally, the conclusions are discussed in \S~\ref{sec:DiscussionFutureResearch}.
\section{Methodology}
\label{sec:Methodology}

This section presents the backbone of the MPC algorithm, with a detailed description of the mathematical tools involved in it. Figure \ref{FIG:Fig1} includes a diagram illustrating all the required steps for its implementation. In addition, algorithm \ref{ALG:Control} is introduced to give more detail on the procedure. 

The main block in the schematics represents the MPC algorithm, following the approach proposed by \citet{Kaiser2018SINDyc}. The main novelty in this module is the robustness enhancement by online filtering with LPR. This is particularly useful when the plant dynamics is modelled with time-delay coordinates or their derivatives. Local polynomial regression is directly applied to past sensor data for online implementation.

The roadmap suggests several necessary steps before implementing the control. First, a training dataset is generated. The dataset consists of the time series of the state dynamics under different exogenous inputs. This data can be collected using various methods, including simulations or experiments. The system state and exogenous inputs should be defined based on the specific system being controlled. Second, a plant model is defined to predict the system behaviour. In this work we use a data-driven nonlinear system identification. 
The final step, before the implementation of the control, focuses on tuning the parameters that define the MPC cost function. Control performance is significantly influenced by their selection. The self-tuning of the hyperparameters is the core of our framework. This tuning is carried out using a BO algorithm.

An important aspect of the proposed method is the need for minimal user interaction. Indeed, the two main inputs are the reference set point (i.e. the control target) and a cost function for the MPC algorithm. The weight of the different contributions in the cost function will be determined in the MPC tuning. Bayesian optimization automatically adjusts to different levels of noise and uncertainties. This greatly enhances the usability and adaptability of the framework to different systems.

\input{FIG_V2/Fig1}

\begin{algorithm}
\caption{Control algorithm \vspace{0.1cm}}
\begin{algorithmic}[1]
\Algphase{\vspace{0.1cm}Training data collection}
\Ensure Training dataset of fluidic pinball forces\\
Choose open loop actuations\\
Resolve fluidic pinball wake using DNS\\
Post process flow fields for $C_d$ and $C_l$ computation
\Algphase{Nonlinear system identification (\S~\ref{subsec:SINDy})}
\Ensure SINDYc fluidic pinball force model: $\dot{a}^k = \boldsymbol{\Theta}(\boldsymbol{a},\boldsymbol{b})\hat{\boldsymbol{\xi}}^k, \,\,\,\, k = 1,\ldots,N_a$ in \eqref{Eqn:SINDYcModel}\\
Construct the matrices $\mathsfbi{A},\dot{\mathsfbi{A}}$ and $\mathsfbi{B}$ in \eqref{Eqn:AAdotBsindy} using training dataset\\
Compute SINDYc active terms of the dynamics:\newline
$\hat{\boldsymbol{\xi}}^k =\underset{ \boldsymbol{\xi^k} }{arg~min} \left\{\frac{1}{2} \norm{\dot{\mathsfbi{A}}_{\bullet,k} -  \boldsymbol{\Theta}(\mathsfbi{A},\mathsfbi{B})\boldsymbol{\xi}^k}^2_2 + \lambda\norm{\boldsymbol{\xi}^k}_1\right\}$ in \eqref{Eqn:OptSindy}
\vspace{0.3cm}
\Algphase{MPC-tuning (\S~\ref{subsec:BayesOpt})}
\Ensure Optimal hyperparameter vector $\boldsymbol{\eta}_{opt}$ solution of \eqref{Eqn:BOproblem}\\
Select different vectors of hyperparameters for seeding
\While {BO stopping criterion not met}\\
Update $\mathcal{J}_{BO}$ sampling points\\
Use GP to update posterior distribution of $\mathcal{J}_{BO}$ given the samples available so far\\
Build and optimise the acquisition function to find the next search point
\EndWhile
\Algphase{MPC application (\S~\ref{subsec:MPC})}
\For {$j = 1, \ldots, n_{BO}$}\\
Measure system state
\If{there is noise in the measure}
\State Use LPR to last sensor data (\S~\ref{subsec:LPR})
\EndIf\\
{Optimization of MPC cost function in \eqref{Eqn:CostFunctionalMPC} using hyperparameters optimised in MPC-tuning}\\
Apply first optimal control component $\boldsymbol{b}_{j+1} = \boldsymbol{b}_{1}^{opt}$\\
Advance fluidic pinball wake of $T_s$ time units
\EndFor
\end{algorithmic}
\label{ALG:Control}
\end{algorithm}

\subsection{Self-tuning model predictive control}
\subsubsection{Model predictive control}
\label{subsec:MPC}
This section describes the MPC implementation. 
It is assumed to have a time-evolving process whose complete description relies on $N_a\geq1$ parameters. These are included in a state vector, denoted at a given instant $t$ as $\boldsymbol{a} \equiv \boldsymbol{a}(t) = \left(a^1(t), \ldots, a^{N_a}(t)\right)'$. The evolution in time of the process can be influenced by the choice of $N_b\geq1$ exogenous parameters. These are included in an input vector $\boldsymbol{b} \equiv\boldsymbol{b}(t) = \left(b^1(t),\ldots, b^{N_b}(t)\right)'$, $\boldsymbol{b} \in \mathcal{B} \subset \mathbb{R}^{N_b}$, where $\mathcal{B}$ is the set of allowable inputs. Denoting the time derivative of the state vector with $\dot{\boldsymbol{a}}$, the system dynamics is described by the following set of equations:
\begin{equation}
\begin{array}{rcl}
    \boldsymbol{\dot{a}} &=& f(\boldsymbol{a},\boldsymbol{b})\\
    \boldsymbol{a}(0) &=& \boldsymbol{a}_0.
    \end{array}
    \label{Eqn:ProcessDynamics}
\end{equation}
Here $\boldsymbol{a}(t_j)=\boldsymbol{a}_j$ and $f$ is the function characterizing the system's temporal evolution. The process is considered as controlled. This means that for each time $t$, the input vector $\boldsymbol{b}(t)$ can be selected in order to manipulate the system dynamics according to a specific objective. More specifically, the aim is to control $N_c\geq1$ features of the dynamical system, included in the vector $\boldsymbol{c} \equiv\boldsymbol{c}(t) = \left(c^1(t), \ldots, c^{N_c}(t)\right)'$. Note that the vector $\boldsymbol{c}$ is dependent on the system state. In this context, it is assumed that the target features are part of the state vector itself, although this may not always be applicable. The objective is to achieve control over the vector $\boldsymbol{c}$ by ensuring that it closely tends to a desired reference $\boldsymbol{c}_{*} \in \mathbb{R}^{N_c}$ over time. Model predictive control can be used for set-point stabilization, trajectory tracking or path following \citep[see][pp. 169-198]{RakovicV2018HandBookMPC}, depending on the choice of $\boldsymbol{c}_{*}$.

\input{FIG_V2/Fig2}

For the purpose of a control application, $\boldsymbol{a}$, $\boldsymbol{b}$ and $\boldsymbol{c}$ are sampled over a discrete-time vector, equispaced with a fixed time interval $T_s$. This discrete-time representation is essential to determine how often the exogenous input is updated. The input is assumed to be constant between consecutive time steps of the control. The implementation of the MPC follows a series of sequential procedures, as can be seen in algorithm \ref{ALG:Control}. An illustration of the process is provided in figure \ref{FIG:Fig2}.

Firstly, the control process starts from a time instant $t_j$, where a measurement $\boldsymbol{s}_j$ of the state vector is available. It is assumed that the entire vector of target features is observed. A conditional prediction of the state vector is obtained by a model of the dynamics. This prediction under a given input sequence, referred to as $\boldsymbol{\hat{a}}_{j+k|j}$, is generated within a prediction window  $t_{j+k}, k=1,\ldots,w_p$. Consequently, a prediction of the target features vector $\boldsymbol{\hat{c}}_{j+k|j}$ is obtained.

The optimal input sequence $\{\boldsymbol{b}_{k}^{opt}\}_{k = 1}^{w_c}$ can be determined in a control window $t_{j+k}, k = 1,\ldots,w_c$, by minimizing a cost function $\mathcal{J}_{MPC} : \mathbb{R}^{N_b} \rightarrow \mathbb{R}^+$. 
A common choice for the cost function is
\begin{equation}
\begin{array}{ccl}
 \mathcal{J}_{MPC}(\boldsymbol{b}) &=& \sum_{k=0}^{w_p} \norm{\boldsymbol{\hat{c}}_{j+k|j} - \boldsymbol{c}_*}^2_{\mathsfbi{Q}} + \\
 & & + \sum_{k=1}^{w_c}(\norm{\boldsymbol{b}_{j+k|j}}^2_{\mathsfbi{R}_b} + \norm{\Delta \boldsymbol{b}_{j+k|j}}^2_{\mathsfbi{R}_{\Delta b}}),
 \label{Eqn:CostFunctionalMPC}
 \end{array}
 \end{equation}
where $\mathsfbi{Q} \in \mathbb{R}^{{N_c} \times {N_c}}$ and $\mathsfbi{R}_b,\mathsfbi{R}_{\Delta b} \in \mathbb{R}^{{N_b} \times {N_b}}$ are positive and semi-positive definite weight matrices, respectively. In addition, $\norm{\boldsymbol{d}}_{\mathsfbi{M}}^2 = \boldsymbol{d}'\mathsfbi{M}\boldsymbol{d}$ represents the weighted norm of a generic vector $\boldsymbol{d}$ with respect to a symmetric and positive definite matrix $\mathsfbi{M}$, where $\boldsymbol{d}'$ denotes the transposition of the vector $\boldsymbol{d}$. The variable $\Delta\boldsymbol{b}_{j+k|j}$ denotes the input variability in time, that is, $\boldsymbol{b}_{j+k|j} - \boldsymbol{b}_{j+k-1|j}$. In the definition of the cost function, the errors in state predictions with respect to the reference trajectory, i.e. $\boldsymbol{\hat{c}}_{j+k|j} - \boldsymbol{c}_*$, are penalised, as well as the actuation cost and variability.
In this paper the aforementioned weight matrices are assumed to be diagonal; thus:
\begin{equation}
    \begin{array}{lllll}
      \mathsfbi{Q}    &=&  \text{diag(} Q^1, \ldots, Q^{N_c}\text{)},\\
      \mathsfbi{R}_b    &=& \text{diag(} R_{b^1}, \ldots, R_{b^{N_b}}\text{)},\\
      \mathsfbi{R}_{\Delta b}   &=&  \text{diag(} R_{\Delta b^1}, \ldots, R_{\Delta b^{N_b}}\text{)}.\\
    \end{array}
    \label{Eqn:WeightMatrices}
\end{equation}

Once the optimization problem in \eqref{Eqn:CostFunctionalMPC} is solved, only the first component of the optimal control sequence, $\boldsymbol{b}_{j+1} = \boldsymbol{b}_{1}^{opt}$, is applied. The optimization is then reinitialised and repeated at each subsequent time step of the control. Generally, the prediction window covers a wider range than the control window ($w_p \geq w_c$). The control vector is considered constant beyond the end of the control window, as discussed in \citet{Kaiser2018SINDyc}. 

Hard constraints on the input vector can readily be incorporated into the optimization process. At each time step, the optimal problem must guarantee that $\boldsymbol{b} \in \mathcal{B}$, where $\mathcal{B}$ is generally determined by the control hardware. The set of allowable inputs is
\begin{align}
\arraycolsep=2pt\def\arraystretch{1.8}
\begin{array}{rcl}
b^k_j &\in& [b^k_{min}\text{ , }b^k_{max}], \\
\Delta b^k_j &\in& [\Delta b^k_{min}, \Delta b^k_{max}], \quad j = 1,\ldots,t \,\, \text{ and }\,\, k = 1,\ldots, N_b ,
\end{array}\label{eq:constraints}
\end{align}
where $b^k_j$ and $\Delta b^k_j$ are the $k^{th}$ control input and input variability component at the 
$j^{th}$ time step. The superscripts \textit{min} and \textit{max} indicate the minimum and maximum admissible values for the control input and its variability between consecutive time steps, respectively. 

A critical issue of this control technique concerns the choice of parameters. Many of these can be selected based on the physics of the system to be controlled or appropriate control hardware limits, such as the actuation constraints or the control time step. The selection of parameters involved in equation \eqref{Eqn:WeightMatrices}, as well as the length of the prediction/control window, is traditionally tuned by trial and error. In \S~\ref{subsec:BayesOpt} we introduce our hyperparameter self-tuning procedure. 
\subsubsection{Nonlinear system identification}
\label{subsec:SINDy}
The optimization loop of MPC requires an accurate plant model for predicting the system's dynamics. The choice of the predictive model typically involves a trade-off between accuracy and computational complexity. In this work we adopt a data-driven sparsity-promoting technique. The framework was illustrated in \citet{Brunton2016Sindy} and later applied in MPC by \citet{Kaiser2018SINDyc} for a variety of nonlinear dynamical systems. It must be remarked that the self-tuning framework proposed here can easily be adapted to accommodate a different plant model, either analytical or data driven, including deep-learning models.

SINDy is a method that identifies a system of ordinary differential equations describing the dynamical system.  
In particular, the version described in this work corresponds to the extension of the SINDy model with exogenous input (SINDYc, \citealp{Brunton2016SINDYc}). 
Considering a system of ordinary differential equations such as the one shown in \eqref{Eqn:ProcessDynamics}, SINDYc derives an analytical expression of $f$ from data. This process requires a dataset comprising the time series of the state vector $\boldsymbol{a}$ and the exogenous input $\boldsymbol{b}$. This method is based on the idea that most physical systems can be characterised by only a few relevant terms, resulting in governing equations that are sparse in a high-dimensional nonlinear function space. The resulting sparse model identification aims to find a balance between model complexity and accuracy, preventing overfitting of the model to the data.

To derive an expression of the function $f$ from data, a discrete sampling is performed on a time vector, yielding $r$ snapshots at time instances $t_i, i = 1, \ldots,r$ for the state vector $\boldsymbol{a}_i = \boldsymbol{a}(t_i)$, its time derivative $\dot{\boldsymbol{a}}_i = \dot{\boldsymbol{a}}(t_i)$ and the input signal $\boldsymbol{b}_i = \boldsymbol{b}(t_i)$.
The data obtained are arranged into three matrices $\mathsfbi{A}\in \mathbb{R}^{r \times N_a}$, $\dot{\mathsfbi{A}}\in \mathbb{R}^{r \times N_a}$ and $\mathsfbi{B}\in \mathbb{R}^{r \times N_b}$:
\begin{equation}
\begin{array}{rcl}
\mathsfbi{A}       &=& \left(\boldsymbol{a}_1\,,\,\ldots\,,\,\boldsymbol{a}_r\right)', \vspace{0.1cm} \\
\dot{\mathsfbi{A}} &=& \left(\boldsymbol{\dot{a}}_1\,,\,\ldots\,,\,\boldsymbol{\dot{a}}_r\right)', \vspace{0.1cm} \\
\mathsfbi{B}       &=& \left(\boldsymbol{b}_1\,,\,\ldots\,,\,\boldsymbol{b}_r\right)'. \vspace{0.1cm} \\
    \label{Eqn:AAdotBsindy}
    \end{array}
\end{equation}

A library of functions $\boldsymbol{\Theta}$ is then set as
\begin{equation}
\begin{array}{cl}
\boldsymbol{\Theta}(\mathsfbi{A},\mathsfbi{B}) =&
(\mathsfbi{1}_r,\,  \mathsfbi{A},\,  \mathsfbi{B},\,
\left((\mathsfbi{A}_{1 \bullet}\otimes\mathsfbi{A}_{1 \bullet})', \ldots,  (\mathsfbi{A}_{r \bullet}\otimes\mathsfbi{A}_{r \bullet})'\right)',\,\\
&\left((\mathsfbi{A}_{1 \bullet}\otimes\mathsfbi{B}_{1 \bullet})', \ldots,  (\mathsfbi{A}_{r \bullet}\otimes\mathsfbi{B}_{r \bullet})'\right)'\\
&\left((\mathsfbi{B}_{1 \bullet}\otimes\mathsfbi{B}_{1 \bullet})', \ldots,  (\mathsfbi{B}_{r \bullet}\otimes\mathsfbi{B}_{r \bullet})'\right)',\ldots),
 \label{Eqn:LibrarySindy}
 \end{array}
\end{equation}
where $\mathsfbi{1}_r$ is a column vector of $r$ ones, $\mathsfbi{A}_{i \bullet}$ denotes the $i^{th}$ row of the matrix $\mathsfbi{A}$ and $\mathsfbi{H}\otimes\mathsfbi{K}$ denotes the Kronecker products of $\mathsfbi{H}$ and $\mathsfbi{K}$.
The choice of functions to be included in the library is typically made by the user. This procedure is often guided by experience and prior knowledge about the dynamics of the system to be controlled. This issue introduces some limitations that are later discussed in \S \ref{sec:DiscussionFutureResearch}.

The system data can be then assumed to be generated from the following model:
\begin{equation}
    \dot{\mathsfbi{A}} = \boldsymbol{\Theta}(\mathsfbi{A},\mathsfbi{B})\boldsymbol{\Xi}.
    \label{SystemSindy}
\end{equation}
$\boldsymbol{\Xi} = \left( Here \boldsymbol{\xi}^1,\ldots, \boldsymbol{\xi}^{N_a} \right)$ represents the matrix whose rows are vectors of coefficients determining which terms of the right-hand side are active in the dynamics of the $k^{th}$ state vector component. This matrix is sparse for many of the dynamical systems considered. It can be obtained by solving the optimization problem
\begin{equation}
    \hat{\boldsymbol{\xi}}^k =\underset{ \boldsymbol{\xi}^k }{arg~min} \left\{\frac{1}{2} \norm{\dot{\mathsfbi{A}}_{\bullet k} -  \boldsymbol{\Theta}(\mathsfbi{A},\mathsfbi{B})\boldsymbol{\xi}^k}^2_2 \,+\, \lambda\norm{\boldsymbol{\xi}^k}_1\right\},
    \label{Eqn:OptSindy}
\end{equation}
where $\lambda$ is called the sparsity-promoting coefficient, $\mathsfbi{A}_{\bullet k}$ denotes the $k^{th}$ column of the matrix $\mathsfbi{A}$. $\norm{\cdot}_1$ and $\norm{\cdot}_2$ are the $L_1$ and $L_2$ norms, respectively. 
It must be remarked that $\lambda$ must be optimised to reach a compromise between parsimony and accuracy. Nonetheless, since this block of the MPC framework can easily be replaced by other strategies, $\lambda$ will not be included in the self-tuning optimization illustrated in \S~\ref{subsec:BayesOpt}.

Finally, the system can be described by
\begin{equation}
    \dot{{a}}^k = \boldsymbol{\Theta}(\boldsymbol{a},\boldsymbol{b})\hat{\boldsymbol{\xi}}^k, \,\,\,\, k = 1,\ldots,N_a,
    \label{Eqn:SINDYcModel}
\end{equation}
where in this case $\boldsymbol{\Theta}(\boldsymbol{a},\boldsymbol{b})$ is a vector that takes into account the same functions included in the library in \eqref{Eqn:LibrarySindy}. The SINDYc-based model is utilised to make predictions of the dynamics involved in the control process starting from a specific state's initial condition. The steps described in this section to obtain the plant model are also condensed in the nonlinear system identification step in algorithm \ref{ALG:Control}.
\subsubsection{Local Polynomial Regression}
\label{subsec:LPR}

Effectively estimating system dynamics is crucial for various control techniques, particularly in MPC, where sensor feedback is utilised to make informed decisions. The accuracy of system dynamics estimation becomes paramount in MPC as it relies on sensor information to predict the behaviour of the controlled system. However, the presence of noise can hinder accurate estimation, necessitating the implementation of noise mitigation techniques.

This challenge falls within the broader context of time series analysis, as sensor outputs represent discrete samples over time of a process variable. To address measurement noise, LPR emerges as a robust, accurate and cost-effective non-parametric smoothing technique. Applying LPR to sensor output data enhances the reliability and accuracy of estimating the current state of the system under varying noise conditions. In this section we provide a succinct overview of the formulation of LPR. The notation convention involves denoting random variables with capital letters, while deterministic values are represented in lowercase.

The objective is to predict or explain the response $S$ (sensor output) using the predictor $T$ (time) from a sample $\left\{(T_j, S_j)\right\}_{j=1}^n$ using the following regression model:
\begin{equation}
    S_j = m(T_j) + \sigma(T_j)\epsilon_j,
    \label{Eqn:NoiseModel}
\end{equation}
Here $m$ is the regression function, $\{\epsilon_j\}_{j=1}^{n}$ are zero mean and unit variance random variables and $\sigma^2$ is the point variance. For simplicity, it is assumed that the model is homoscedastic, i.e. that the variance is constant. From a practical point of view, $\sigma$ also represents the measurement of noise intensity.

The reason why LPR is used in this context is that it allows for a fast joint estimation of the regression function and its derivatives without making any prior. By making only a few regularity assumptions, it is characterised by a good flexibility that allows us to model complex relations beyond a parametric form. The working principle of the LPR is to approximate the regression function locally by a polynomial of order $p$, performing a weighted least squares regression using data only around the point of interest.

It is assumed that the regression functions admit derivatives up to the order $p+1$ and, thus, that it can be expanded in a Taylor's series. For close time instants $t$ and $T_j$, the unknown regression function can be approximated by a polynomial of order $p$, then
\begin{equation}
    m(T_j) \approx \sum_{i=0}^p \frac{m^{(i)}(t)}{i!}(T_j-t)^i \equiv \sum_{i=0}^p \beta_i(T_j-t)^i,
    \end{equation}
The vector of parameters $\boldsymbol{\beta} = (\beta_0,\ldots,\beta_p)'$ can be estimated by solving the minimization problem
\begin{equation}
\boldsymbol{\hat{\beta}} = \underset{\boldsymbol{\beta}}{\text{ arg min }} \sum_{j=1}^{n} \{S_j - \sum_{i=0}^p \beta_i(T_j-t)^i \}^2 K_h(T_j-t),
\label{Eqn:LPRminprob}
\end{equation}
where $h$ is the bandwidth determining the size of the local neighbourhood (also called smoothing parameter) and $K_h(t) = \frac{1}{h}K(\frac{t}{h})$ with $K$ a kernel function assigning the weights to each observation. Once the vector $\hat{\boldsymbol{\beta}}$ has been obtained, an estimator of the $q^{th}$ derivative of the regression function is
\begin{equation}
    \hat{m}^{(q)}(t) = q!\hat{\beta}_q.
\end{equation}

The solution of the optimization in \eqref{Eqn:LPRminprob} is simplified when the methodology is presented in matrix form. To that end, $\mathsfbi{T}\in \mathbb{R}^{n\times(p+1)}$ is the design matrix
\begin{equation}
\mathsfbi{T} = 
\begin{pmatrix}
   1        &   (T_1-t)  &   \ldots   &    (T_1-t)^p    \\
    \vdots  &   \vdots   &            &     \vdots       \\
    1       &   (T_{n}-t)  &   \ldots   &    (T_{n}-t)^p     
\end{pmatrix}
\end{equation}
while the vector of responses is $\boldsymbol{S} = (S_1,\ldots,S_{n})'$ and $\mathsfbi{W} = \text{diag}( K_h(T_1-t), \ldots, K_h(T_{n}-t))$.
According to this notation and assuming the invertibility of $\mathsfbi{T}'\mathsfbi{W}\mathsfbi{T}$, the weighted least squares solution of the minimization problem expressed in \eqref{Eqn:LPRminprob} is
\begin{equation}
\boldsymbol{\hat{\beta}} = (\mathsfbi{T}'\mathsfbi{W}\mathsfbi{T})^{-1}\mathsfbi{T}'\mathsfbi{W}\boldsymbol{S}
\end{equation}
and, thus, the estimator of the local polynomial is
\begin{equation}
\hat{m}(t) = \boldsymbol{e}_1'(\mathsfbi{T}'\mathsfbi{W}\mathsfbi{T})^{-1}\mathsfbi{T}'\mathsfbi{W}\boldsymbol{S} = \sum_{j = 1}^{n} W_j^p(t)S_j,
\end{equation}
where $\boldsymbol{e}_k\in\mathbb{R}^{p+1}$ a vector having $1$ in the $k^{th}$ entry and zero elsewhere and $W_j^p(t) = \boldsymbol{e}_1'(\mathsfbi{T}'\mathsfbi{W}\mathsfbi{T})^{-1}\mathsfbi{T}'\mathsfbi{W}\boldsymbol{e}_j$.

The critical element that determines the degree of smoothing in LPR is the size of the local neighbourhood. The selection of $h$ determines the accuracy of the estimation of the regression function: too large values lead to a large bias in the regression, while too small values increase the variance. Its choice is generically the result of a trade-off between the variance and the estimation bias of the regression function. There are several methods to select the bandwidth to be used for LPR. One possible approach is to choose between a global bandwidth, which is common to the entire domain and is optimal for the entire range of data, or a local bandwidth that depends on the covariate and is therefore optimal at each point of the regression function estimation. The latter would allow for more flexibility in estimating inhomogeneous regression functions. However, in the proposed framework, a global bandwidth was chosen due to its simplicity and, more importantly, its reduced computational cost. In the presented approach, the global bandwidth is chosen using the leave-one-out cross-validation methodology. Specifically, this parameter corresponds to the one that minimises the following expression:
\begin{equation}
    \sum_{j = 1}^{n} \left(S_j - \hat{m}_{h, -j}(T_j)\right)^2.
\end{equation}
Here $\hat{m}_{h,-j}(T_j)$ denotes the estimation of the regression function while excluding the $j^{th}$ term. For further information regarding the optimal bandwidth selection, see \citet{Wand1994KernelSmoothing} and \citet{fan1996LPRapplications}. Additionally, there exist other methodologies that utilise machine-learning techniques to select the local bandwidth such as in \citet{Giordano2008NN4BandSelectionLLR} where neural networks are used.

Once the smoothing parameter has been set, it remains to choose the weighting function and the degree of the local polynomial, although these two have a minor influence on the performance of the LPR estimation. A common choice for the first is the Epanechnikov kernel. As for the degree of the local polynomial, there is a general pattern of increasing variability according to which, to estimate $m^{(q)}(t)$, the lowest odd order is recommended, i.e. $p = q + 1$ or occasionally $p = q + 3$ \citep{fan1996LPRapplications}.

It is also worth noting that due to asymmetric estimation within the control algorithm, boundary effects arise, leading to a bias in the estimation at the edge. These effects, discussed in more detail in \citet{fan1996LPRapplications}, disappear when using local linear regression ($p = 1$).
\subsubsection{Hyperparameter automatic tuning with BO}
\label{subsec:BayesOpt}

Model predictive control relies on a specific set of hyperparameters to precisely define the cost function in \eqref{Eqn:CostFunctionalMPC} for the selection of the optimal action over the control window. A new functional, based on the global performance of the control, is defined. Bayesian optimization is employed to find the hyperparameter vector that maximises the control performance. By adopting this approach, a more efficient and effective MPC framework may be achieved. The hyperparameters are indeed adapted to different conditions of measurement noise and/or model uncertainty by the BO process. This proposed framework is practically independent of the user.

The parameters under consideration include \textit{i)} the components of the weight matrices, as presented in (\ref{Eqn:WeightMatrices}), which penalise errors of the state vector with respect to the control target trajectories, input cost and input time variability; and \textit{ii)} the length of the control/prediction windows, assumed equal for simplicity.
All aforementioned control parameters are included in a single vector denoted as $\boldsymbol{\eta} \in \mathbb{R}^{N_{\boldsymbol{\eta}}}$, where $N_{\boldsymbol{\eta}} = 2 N_b + N_c + 1$. A further reduction in the number of control parameters can also be considered by imposing that the components of $\mathsfbi{R}_b$ and $\mathsfbi{R}_{\Delta b}$ related to the rear cylinders of the fluidic pinball are identical under a flow symmetry argument, as in \citet{Bieker2019DeepMPC}. 
Therefore, in this case it is stated that ${R}_{b^2} = {R}_{b^3} = {R}_{b^{2,3}}$ and ${R}_{\Delta b^2} = {R}_{\Delta b^3} = {R}_{\Delta b^{2,3}}$ and the total number of parameters is reduced to $N_{\boldsymbol{\eta}} = 2(N_b-1) + N_c + 1$. Section \ref{sec:Results} also provides several results that justify this choice. 

Note that control results depend on the selection of the vector $\boldsymbol{\eta}$. Consequently, an offline optimization process can be implemented to select the optimal value of $\boldsymbol{\eta}$, which maximises the control performance.
Consider using a specific realization of the hyperparameter vector $\boldsymbol{\eta}$  to control the system. The state vector measure is indirectly dependent on the choice of hyperparameter vector and is denoted here by $\tilde{\boldsymbol{s}}(\boldsymbol{\eta})$. By running the control on a discrete-time vector of $n_{BO}$ time steps, the cost function can be defined as
\begin{equation}
    \mathcal{J}_{BO}(\boldsymbol{\eta}) = \frac{1}{n_{BO}} \sum_{k=1}^{N_c}\sum_{j=1}^{n_{BO}}\left(\tilde{s}^k_j(\boldsymbol{\eta}) - c^k_*\right)^2.
    \label{Eqn:BOcost}
\end{equation}
Note that the user selects the target to be optimised, specified by the cost function $\mathcal{J}_{BO}$. In addition, in absence of measurement noise, $\tilde{\boldsymbol{s}}(\boldsymbol{\eta})$ is the ideal measure of the target feature vector. Otherwise, the LPR technique is applied to the state vector measure. In this study the parameters are optimised to minimise the quadratic error between the controlled variables and the user-defined target.
Thus, the optimal hyperparameter vector can be obtained by solving the problem
\begin{equation}
\begin{array}{rl}
   \boldsymbol{\eta}_{opt} = \underset{\boldsymbol{\eta} \in H}{\text{arg min}} & \mathcal{J}_{BO} (\boldsymbol{\eta}),\\
   \label{Eqn:BOproblem}
    \end{array}
\end{equation}
where $H \subset \mathbb{R}^{N_{\boldsymbol{\eta}}}$ is the search domain. Solving the optimization in \eqref{Eqn:BOproblem} is computationally costly due to the expensive sampling of the black-box cost function $\mathcal{J}_{BO}$. Each sample of $\mathcal{J}_{BO}$ is obtained via application of the MPC over $n_{BO}$ time steps. In this framework, BO serves as an efficient method to address this problem, offering an algorithm to search the minimum of the function with a high guarantee of avoiding local minima and characterised by rapid convergence. Indeed, BO has shown to be an efficient strategy particularly when $N_{\boldsymbol{\eta}} \leq 20$ and the search domain $H$ is a hyper-rectangle, that is $H = \{\boldsymbol{\eta}\in \mathbb{R}^{N_{\boldsymbol{\eta}}}| \eta^i \in [\eta^i_{min},\eta^i_{max}]\subset \mathbb{R}\,, \eta^i_{min}<\eta^i_{max}\,,i = 1, \ldots , N_{\boldsymbol{\eta}} \}$, where $\boldsymbol{\eta}_{min}$ and $\boldsymbol{\eta}_{max}$ are the lower and upper bound vectors of the hyper-rectangle, respectively.

In order to find the minimum of the unknown objective function, BO employs an iterative approach, as shown in the MPC-tuning step of algorithm \ref{ALG:Control}. It utilises a probabilistic model, typically a Gaussian process (GP), to estimate the behaviour of the objective function. At each iteration, the probabilistic model incorporates available data points to make predictions about the function behaviour at unexplored points in the search space. Simultaneously, an acquisition process guides the samplings by suggesting the optimal locations in order to discover the minimum point. A specific function (called the acquisition function) is set to balance exploration, by directing attention to less-explored areas of the search domain, and exploitation, by concentrating on regions near potential minimum points. Finally, the BO iterations continue until a stopping criterion is met, such as reaching a maximum number of iterations or achieving convergence in the search for the minimum. Thus denoting as $\alpha(\boldsymbol{\eta})$ the acquisition function selected, the point where to sample next in the iterative approach, that is, $\boldsymbol{\eta}^+$, can be obtained by solving
\begin{equation}
    \boldsymbol{\eta}^+ = \underset{\boldsymbol{\eta} \in H}{\text{arg max}}\,~ \alpha(\boldsymbol{\eta}).
\end{equation}
The expected improvement is used as an acquisition function \citep{Snoek2012BayesOpt}, which evaluates the potential improvement over the current best solution.

The subsequent discussion will centre on the probabilistic model utilised in BO. Specifically, a prior distribution, which corresponds to a multivariate Gaussian distribution, is employed. Consider the situation where problem \eqref{Eqn:BOproblem} is tackled using BO. A GP regression is required for the functional $\mathcal{J}_{BO}$ based on the available observations of the function up to the given iteration. 

Since $\mathcal{J}_{BO}$ is a Gaussian process, for any collection of $D$ points, included in $\boldsymbol{\Xi} = \left(\boldsymbol{\eta}_1, \ldots, \boldsymbol{\eta}_D\right)'$, then the vector of function samplings at these points, denoted as $\boldsymbol{J} = \left(\mathcal{J}_{BO}(\boldsymbol{\eta}_1),\ldots ,\mathcal{J}_{BO}(\boldsymbol{\eta}_D)\right)'$ is multivariate-Gaussian distributed, then
\begin{equation}
    \boldsymbol{J} \sim \mathcal{N}_D(\boldsymbol{\mu}, \mathsfbi{K})
\end{equation}
where $\boldsymbol{\mu} = \left(\mu(\boldsymbol{\eta}_1),\ldots, \mu(\boldsymbol{\eta}_D)\right)'$ is the mean vector and $\mathsfbi{K}$ the covariance matrix whose $ij^{th}$ component is $K_{ij} = k(\boldsymbol{\eta}_i,\boldsymbol{\eta}_j)$, with $\mu$ a mean function and $k$ a positive definite kernel function. For simplicity, the mean function is assumed to be null.

In order to make a prediction of the value of the unknown function at a new point of interest $\boldsymbol{\eta}_*$, denoted as $\mathcal{J}_{BO_*}$, conditioned on the values of the function already observed, and included in the vector $\boldsymbol{J}$, the joint multivariate Gaussian distribution can be considered:
\begin{equation}
    \begin{pmatrix}
        \boldsymbol{J}\\
        \mathcal{J}_{BO_*}
    \end{pmatrix}
    \sim \mathcal{N}_{D+1}\left(
        \boldsymbol{0},
        \begin{pmatrix}
        \mathsfbi{K} & \boldsymbol{k}_*\\
        \boldsymbol{k}'_* & k_{**}
    \end{pmatrix}
    \right).
\end{equation}
Here $k_{**} = k(\boldsymbol{\eta}^*,\boldsymbol{\eta}^*)$ and the vector $\boldsymbol{k}_* = \left(k(\boldsymbol{\eta}_1, \boldsymbol{\eta}_*), \ldots,k(\boldsymbol{\eta}_D, \boldsymbol{\eta}_*)\right)'$. This equation describes how the samples $\boldsymbol{J}$ at the locations $\boldsymbol{\Xi}$ correlate with the sample of interest $\mathcal{J}_{BO_*}$, whose conditional distribution is
\begin{equation}
\mathcal{J}_{BO_*}|\boldsymbol{\eta}_*,\boldsymbol{\Xi},\boldsymbol{J} \sim \mathcal{N}\left(\mu_* \,,\,  \sigma_*^2\right)
\label{Eqn:PosteriorGP}
\end{equation}
with mean $\mu_* = \boldsymbol{k}'_*\boldsymbol{K}^{-1}\boldsymbol{J}$ and variance $\sigma_*^2 = \left(k_{**} - \boldsymbol{k}'_*\boldsymbol{K}^{-1}\boldsymbol{k}_*\right)^2$. 
Equation \eqref{Eqn:PosteriorGP} provides the posterior distribution of the unknown function at the new point where the sample has to be performed. It then furnishes the surrogate model used to describe the function $\mathcal{J}_{BO}$ in the domain for the search of the minimum point.
\subsection{The fluidic pinball}
\label{subsec:TheFluidicPinball}
The proposed framework is tested on the control of the two-dimensional viscous incompressible flow around a three-cylinder configuration, commonly referred to as fluidic pinball \citep{Pastur2018ROMpinball,Deng2020ROMpinball,Deng2022ClustHierNetpinball}. The fluidic pinball was chosen because it represents a suitable multiple-input-multiple-output system benchmark for flow controllers.

A representation of the fluidic pinball can be seen in figure \ref{FIG:Fig3}. The three cylinders have identical diameters $D = 2R$, and their geometric centers are placed at the vertices of an equilateral triangle of side $3R$. The centres are symmetrically positioned with respect to the direction of the main flow. The leftmost vertex of the triangle points upstream while the rightmost side is orthogonal to the flow direction. The free stream has a constant velocity equal to $U_\infty$.  The cylinders of the fluidic pinball, denoted here with $1$ (front), $2$ (top) and $3$ (bottom), can rotate independently around their axes (orthogonal to the plane of the flow)  with tangential velocity $b^1$, $b^2$ and $b^3$, respectively. 

\input{FIG_V2/Fig3}

The dynamics of the wake past the fluidic pinball is obtained through a two-dimensional direct numerical simulation (DNS) with the code developed by \citet{Noack2017pinballToolkit}. The flow is described in a Cartesian reference system in which the $x$ and $y$ axes are in the streamwise and crosswise directions, respectively. The centre of the Cartesian reference system coincides with the mid-point of the rightmost bottom and top cylinder and the computational domain, that is, $\Omega$, is bounded in $(-5D, 20D) \times (-5D, 5D)$.
The position in the reference system is then described by the vector $\boldsymbol{x} = (x, y) = x \boldsymbol{e_1} + y \boldsymbol{e_2}$, where $\boldsymbol{e_1}$ and $\boldsymbol{e_2}$ are respectively the unit vectors in the directions of the $x$ and $y$ axes and the velocity vector is assumed to be $\boldsymbol{u} = (u, v)$. The constant density is denoted by $\rho$, the kinematic viscosity of the fluid by $\nu$. All quantities used in the discussion are assumed non-dimensionalised with cylinder diameter, free-stream velocity, and fluid density.
The Reynolds number is defined as $\Rey_D = \frac{U_\infty D}{\nu}$. A value of $\Rey_D = 150$ is adopted, which is sufficiently large to ensure a chaotic behaviour, although still laminar. The two-dimensional solver has already been used in previous work at this same $\Rey_D$ \citep{Wang2023cluster}.
The boundary conditions comprise a far-field condition ($\boldsymbol{u} = U_\infty \boldsymbol{e_1}$) in the upper and lower edges, a stress-free one in the outflow edge, and a no-slip condition on the cylinder walls, which in the absence of forcing becomes $\boldsymbol{u} = 0$. 

The DNS allows forcing by independent rotation of the cylinders. To this purpose, a velocity with module $|b^i|$ is imposed at the cylinder surface. Positive values of $b^i$ are associated with counterclockwise rotations of the cylinders.
In the remainder of the paper, a reference time scale is set as the convective unit ($c.u.$), i.e. the time scale based on the freestream velocity and the cylinder diameter. The lift coefficient is defined as $C_l = \frac{2F_l}{\rho  U_\infty^2 D}$, where $F_l$ is the total lift force applied to the cylinders in the direction of the $y$ axis and the same quantity is applied in the scaling of $F_d$, force applied to the cylinders in the direction of the $x$ axis, to obtain the total drag coefficient $C_d$. The DNS uses a grid of 8633 vertices and 4225 triangles accounting for both accuracy and computational speed. A preliminary grid convergence study at $\Rey_D=150$ identified this as sufficient resolution for errors of up to $3\%$ in the free case and $\approx 4\%$ in the actuated case in terms of drag and lift.
\subsection{The control approach}
The aim of the control is to achieve a reduction in the overall drag coefficient of the fluidic pinball, while also controlling the lift coefficient so that the latter has reduced oscillations with a zero average value. The vector of target features is therefore composed only of the total lift and drag coefficients, $\boldsymbol{c} = (C_d, C_l)'$.
The target vector is thus composed of null components $\boldsymbol{c_*} = (0,0)'$. Since the cost function in \eqref{Eqn:CostFunctionalMPC} is quadratic, a null target vector will penalise both mean values of $C_l$ and $C_d$ and their oscillations. 

To this purpose,  the tangential velocities of the three cylinders are tuned respecting the implementation limits, here chosen equal to $b^i \in [-1, 1]$ and $\Delta b^i \in [-4, 4]$ $, \text{for all} i = 1,2,3$.
The model plant has a state vector composed by $C_d$ and $C_l$ and their time derivatives, respectively $\dot{C_d}$ and $\dot{C_l}$ (i.e. $\boldsymbol{a} = (C_d, C_l, \dot{C_d}, \dot{C_l})'$). A state vector based solely on global variables does not require adding intrusive probes for flow estimation.
Similar state vector choices that incorporate aerodynamic force coefficients and temporal derivatives are made in \cite{Nair2019ClusterFeedbControl} and \cite{Loiseau2018SparseredordModelling}. While this approach is often effective in separated flows, it might be challenged at higher $\Rey$ flows and of unfeasible application in other flow configurations.

Feedback to the control involves measurement of drag and lift forces, so at each time instant $t_j$, it is considered $\boldsymbol{s}_j = \boldsymbol{c}_j$. This approach also facilitates a reduction in the complexity of the predictive model, thereby speeding up the control process.
Indeed, a compact state vector is particularly desirable to reduce the computational cost of the iterative optimization problem over receding horizons of MPC.

In the present work, noise in lift and drag measurement is also considered. Under this assumption, the model in equation \eqref{Eqn:NoiseModel} is applied. The noise intensity $\sigma$ is assumed to be constant over time and given as a percentage of the full-scale measured drag and lift coefficients in conditions without actuation. Therefore, in the case of measurement noise in the sensors, the response variable to which the LPR is applied corresponds to the measurements of the drag and lift coefficients of the fluidic pinball. The LPR enables us to obtain estimates of their regression function in output from the sensors but also of their time derivatives, allowing us to use this information as control feedback.

The MPC-optimization problem is solved every control time step (so every $T_s$) to update the exogenous control input. During the time between two consecutive samples, the control input to the system remains constant. Thus, the sampling time step should be chosen small enough to ensure a good closed-loop performance, but not too small to avoid an excessive computational cost. In this work, it is set at $T_s = 0.5\, c.u.$, i.e. sufficiently small compared with the shedding period of the fluidic pinball wake (denoted as $T_{sh}$) and not too high to affect control performance. The reason why $T_s$ was not included in the control tuning concerns the difficulty of defining the search domain for realistic applications. Imposing bounds on this parameter requires an estimate of the computational cost of the MPC, and this procedure is postponed to future experimental applications. The method used to optimise the control action in the MPC framework is sequential quadratic programming with constraints. The optimization was carried out with a built-in function of MATLAB. The stop criteria are set at a maximum number of iterations of $500$ and a step tolerance of $1e{-6}$.

At each measurement taken during the control process, in the presence of noise in the $C_l$ and $C_d$ sensors, the LPR technique is applied to a time series of outputs of length corresponding to the characteristic shedding period of the fluidic pinball wake, as observed in \S~\ref{sec:Results}. Local polynomial regression acts asymmetrically, i.e. only past output sensor data are available. 

Parameter tuning is performed by evaluating the cost function $\mathcal{J}_{BO}$ over a time history of the controlled variables of one shedding period of the fluidic pinball. The transient phase is excluded. Finally, as a stopping criterion for the search for the optimal hyperparameter vector, a maximum number of 100 iterations in BO was set. This proved sufficient to reach convergence in the tuning process, as shown in \S~\ref{SubSec:BOresults}.
\section{Results}
\label{sec:Results}
In this section the results of the current work are presented. Subsection \ref{SubSec:PredictionPerformance} outlines the prediction performance of the SINDYc-based force model. Furthermore, the prediction performance enhancement when using LPR in the presence of measurement noise is illustrated. In \S~\ref{SubSec:BOresults} the effectiveness of hyperparameters tuning in the different control scenarios is examined. Lastly, in \S~\ref{SubSec:ControlResults} the outcomes of applying the MPC algorithm to the fluidic pinball wake for drag reduction purposes under ideal measurement conditions and in the presence of sensor noise are presented. An additional discussion is provided to justify the choice of symmetry in the parameters of the rear cylinders of the fluidic pinball.
\subsection{Predictive model}
\label{SubSec:PredictionPerformance}

This subsection presents the performance evaluation of the predictive model employed in the control framework.
SINDYc is applied on a training dataset consisting of a time series of the system state, along with predetermined actuation laws.
The chosen open-loop laws are composed of step signals for the three-cylinder velocities $b^i$, encompassing all possible combinations of the velocities of the three cylinders within control constraints. Specifically, velocities ranging from $-1$ to $1$ with an increment of $0.5$ are explored, resulting in a total of $5^3$ combinations. For each explored velocity combination, the actuation steps are long enough ($\approx 55\, c.u.$) to allow reaching the respective steady state. The total length of the generated dataset is $6950 \, c.u.$. The actuation time series is subsequently smoothed to include the transient dynamics in the force model. It must be noted that this approach is effective for this test case whose dynamics evolve on a clearly defined low-dimensional attractor. Open-loop training for plant identification has been shown effective in separated flows (see, e.g.  \citealp{kaiser2017cluster,Bieker2019DeepMPC}). More complex nonlinear dynamical systems might require different strategies for adequate training of the modelling plant.

Additionally, a library of polynomial functions up to the second order is chosen for the sparse regression with SINDYc. 

Figure \ref{FIG:Fig4} shows the prediction performance of the plant model. 
The presented results were generated by performing predictions on a testing dataset with a preassigned smooth law of the rotational speeds of the three cylinders. Predictions of $4\, c.u.$ using initial conditions at random points of the validation dataset, and for a statistically significant number of times were performed. The resulting time series of each prediction were then used to calculate their respective errors ($\hat{e}_{C_d}$ for $C_d$ and $\hat{e}_{C_l}$ for $C_l$). The plot illustrates the trend of the average and the confidence region for $\sigma$ of the prediction errors, respectively normalised with the standard deviation of $C_d$ and $C_l$ in the free response, here denoted as $sd(C_{d_0})$ and $sd(C_{l_0})$, respectively.

\input{FIG_V2/Fig4}

The panels on the left show the prediction performance calculated from an initial condition obtained in the absence of measurement noise in the sensors. The average value of the normalised error distribution maintains values close to zero for any length of the prediction window analysed. Furthermore, for short prediction lengths (up to $\approx3 \, c.u.$) the confidence bounds for $\sigma$ are still confined within a narrow region. In this context, as they are not directly measurable, the time derivatives of the aerodynamic coefficients are calculated using finite differences. This approach, however, is highly sensitive to measurement noise. The non-parametric smoothing technique LPR allows joint estimation of the output data from the sensors and their respective time derivatives. The panels on the right show the benefits brought by initial-condition setting with LPR to the prediction performance in the presence of a sensor measurement noise of $1\%$. These plots show a comparison of the statistical error distributions for two distinct cases: in one the initial condition corresponds to the LPR estimation of the data with noise, in the other, it is not used. It is of considerable interest to observe that the prediction with LPR performs similarly to the case without noise. The figure discussed above could be used to choose the MPC prediction window length as it provides an idea of the uncertainty propagation in the prediction horizon. However, since this parameter has a significant effect on control results in terms of performance and computational cost, it is left to automatic selection through BOn, so as to also take into account the effects of measurement noise in its selection.

In addition, figure \ref{FIG:Fig5} provides further insights into the advantages of LPR. In this case the estimation of $C_d$ and $\dot{C_d}$ is tested on a case with external forcing and under increasing noise level. The trend in the LPR estimation error of both $C_d$ and $\dot{C_d}$ is illustrated as the noise increases. The estimation errors are evaluated using a root-mean-square error (RMSE) of the estimate compared with the ideal data without noise.

\input{FIG_V2/Fig5}
The results demonstrate the effectiveness of LPR in estimating $C_d$ and $\dot{C_d}$ with high accuracy even in the presence of high levels of noise. Thus, the benefits brought by LPR enable the control feasibility even in high noise scenarios.
\subsection{Bayesian optimization for tuning control parameters}
\label{SubSec:BOresults}
This subsection presents the results of the hyperparameters tuning using the BO algorithm. Various control scenarios are analysed, including clean, low and high noise level cases.

The set of hyperparameters to be optimised is composed of the elements of the weight matrices in \eqref{Eqn:WeightMatrices} and the length of the control/prediction window, used in the definition of the cost functional in \eqref{Eqn:CostFunctionalMPC}. Specifically, the components of the matrix $\mathsfbi{Q}$ are optimised within the interval $[0.1, 5]$, the terms of $\mathsfbi{R}_{b}$ and $\mathsfbi{R}_{\Delta b}$ in $[0.1, 10]$ and $w_p$ is optimised within $[1, 4]$.

Figure \ref{FIG:Fig6} displays the trend of the minimum observed $\mathcal{J}_{BO}$ samplings as the tuning process iterations progress for some of the analysed control cases. The results indicate that the optimization function ($\mathcal{J}_{BO}$) reaches convergence in less than 30 iterations for almost all cases.
\input{FIG_V2/Fig6}
The tuned control parameters of each case are presented in table \ref{TAB:TuningResults}. The table also provides the average values of $C_d$ and $C_l$ ($\bar{C_d}$ and $\bar{C_l}$), as well as their corresponding standard deviations ($sd(C_d)$ and $sd(C_l)$), during the control phase. These values are computed over a shedding time interval. 

It must be remarked that the final set of hyperparameters for each case is rundependent. This is mostly to be ascribed to the weak dependence of $\mathcal{J}_{BO}$ to some of the imposed constraints in the MPC loss function. Nonetheless, the differences in terms of drag reduction have been observed to be smaller than $\pm 1.1\%$ in different runs.

Figure \ref{FIG:Fig7} presents a plot of the cost function $\mathcal{J}_{BO}$ and its contributions due to the control of $C_d$ (denoted as $\mathcal{J}_{BO}^1$) and $C_l$ (denoted as $\mathcal{J}_{BO}^2$), varying $Q^1$ and $Q^2$, and then $R_{b^1}$ and $R_{b^2} = R_{b^3} = R_{b^{2,3}}$ under $1\%$ measurement noise. The two contributions of $\mathcal{J}_{BO}$ can be obtained by considering the cost functional in equation \eqref{Eqn:BOcost} and selecting only the component for $k = 1$ (to obtain $\mathcal{J}_{BO}^1$) and, for $k = 2$, to obtain $\mathcal{J}_{BO}^2$. The presented cost function plot is useful for interpreting the convergence parameters of the optimization process according to BO.
The term $\mathcal{J}_{BO}^{2}$ has a minor influence on the optimization process, being approximately two orders of magnitude smaller than $\mathcal{J}_{BO}^1$. Therefore, the optimization process is mainly driven by the influence of the control parameters on reducing the $C_d$ of the fluidic pinball.

As $Q^1$ increases, both $\mathcal{J}_{BO}$ and $\mathcal{J}_{BO}^1$ functions decrease sharply at first and then reach a plateau. Conversely, $\mathcal{J}_{BO}^2$ shows the opposite behaviour. A good $C_l$ control would be achieved with a low $Q^1$ and high $Q^2$, but this would lead to a poor drag reduction. On the other hand, the choice of $Q^2$ has little impact on control performance, as the cost function $\mathcal{J}_{BO}$ remains nearly constant with its variation.

By observing table \ref{TAB:TuningResults} and considering the behaviour of the cost function, it is justified why almost all proposed control cases have $Q^1$ very close to the maximum achievable value, while there is greater variability in the selection of $Q^2$.

Similarly, due to the fact that the drag reduction drives the optimization process, the parameter that has the greatest influence on $\mathcal{J}_{BO}$ is $R_{b^{2,3}}$. A lower value of $R_{b^{2,3}}$ would assign less weight to the actuation cost of the rear cylinders of the fluidic pinball, allowing for greater rotation intensity. As will be explained in \S~\ref{SubSec:NoNoiseRes}, the main contributors to drag reduction are indeed the rear cylinders. On the other hand, the parameter $R_{b^1}$ has little impact on the overall control performance since the front cylinder is responsible for the stabilization of the lift coefficient. These latter considerations justify that the optimum $R_{b^{2,3}}$ is achieved at low values, close to the lower possible limit, while greater variability is observed in the optimization of the $R_{b^1}$ parameter.

\begin{table}
  \centering
  {\renewcommand\arraystretch{1.8}
  \begin{adjustbox}{max width=\textwidth}
    \begin{tabular}{cccccccccccc}
    Noise       & LPR  & $diag(\mathsfbi{Q}$) & diag($\mathsfbi{R}_b$) & diag($\mathsfbi{R}_{\Delta b}$) & $w_p$ $(c.u.)$ & $\bar{C_d}$ & $sd(C_d)$ & $\bar{C_l}$ & $sd(C_l)$ \\
     0\% & no & (1.000, 1.000) & (1.000, 1.000, 1.000) & (1.000, 1.000, 1.000) & 3.0 & 3.0598 & 0.0956 & -0.0530 & 0.1335\\
    $0\%$   & no   & (4.703, 0.206) & (6.334, 3.735, 1.147)$^*$ & (1.754, 0.176, 0.768)$^*$ & 3.4 & 1.9685 & 0.0099 & -0.2278  & 0.0687 \\
    $0\%$   & no   & (4.881, 0.435) & (5.133, 0.903, 0.903)     & (8.409, 0.193, 0.193)     & 3.2 & 1.9633 & 0.0094 &  0.0022  & 0.0604 \\
    $0.5\%$ & no   & (4.886, 0.191) & (7.435, 0.514, 0.514)     & (3.961, 0.465, 0.465)     & 3.8 & 1.9373 & 0.0200 &  0.0042  & 0.0863 \\
    $0.5\%$ & yes  & (3.836, 2.283) & (9.776, 0.337, 0.337)     & (0.367, 1.465, 1.465)     & 3.9 & 1.9294 & 0.0099 &  0.0172  & 0.0909 \\
    $1\%$   & yes  & (4.938, 1.753) & (5.960, 0.170, 0.170)     & (1.123, 2.521, 2.521)     & 4.0 & 1.9284 & 0.0091 &  0.0103  & 0.0914 \\
    $5\%$   & yes  & (4.659, 1.007) & (9.556, 0.547, 0.547)     & (0.135, 1.044, 1.044)     & 3.0 & 1.9371 & 0.0100 &  0.0110  & 0.0970 \\
    \end{tabular}
    \end{adjustbox}}
      \caption{\justifying Results of the control tuning through Bayesian optimization. A comparison is provided for all the analyzed simulation cases, with a particular focus on the tuning effects due to the intensity of the sensor measurement noise, the use of the LPR technique in the sensor outputs, and the imposition of symmetry in the control parameters. The asterisk ($^*$) denotes control cases in which the tuning is performed on all control parameters, without imposing the symmetry condition on the parameters of the rear cylinders of the fluidic pinball. The table also provides the case with all parameters equal to $1$ (except for the prediction/control window set at $3 c.u.$)}
  \label{TAB:TuningResults}
\end{table}

\input{FIG_V2/Fig7.tex}
\subsection{Application of the MPC algorithm}
\label{SubSec:ControlResults}
This subsection shows the results of applying the control algorithm to the wake of the fluidic pinball. It is specified that all results reported from now on are obtained by imposing symmetry in the parameters of the rear cylinders.
\subsubsection{Control simulations without measurement noise}
\label{SubSec:NoNoiseRes}
Before discussing the effects of MPC, a brief description of the behaviour of the unforced wake of the fluidic pinball in the laminar chaotic regime, characteristic of the chosen Reynolds number ($\Rey_D = 150$), is presented.

Time histories of lift (plot a) and drag (plot b) coefficients for $500\, c.u.$ are reported in figure \ref{FIG:Fig8}. The figure also includes a power spectral density (PSD) of the lift coefficient (plot c), with the Strouhal number $\mathit{St}_D = \frac{fD}{U_\infty}$ related to the diameter of the three cylinders, and a representation of the trajectory in the time-delayed embedding space of the force coefficients (plot d). In the chaotic regime of the fluidic pinball, the main peak in the PSD of $C_l$ is notably broad. However, a predominant one is found corresponding to $\mathit{St}_D = 0.148$, associated with vortex shedding (with a shedding period $T_{sh} = 6.76\, c.u.$). The curve also shows a second peak associated with a lower Strouhal number ($\mathit{St}_D = 0.013$) due to resonance in the flow field related to the finite size of the domain, as also observed in \citet{Deng2020ROMpinball}.

The condition of flow symmetry ($\bar{C_l}\approx 0$) is recovered in the chaotic regime, unlike the lower-Reynolds-number regimes which exhibit non-symmetric wakes \citep{Deng2020ROMpinball}. In addition, $C_l$ standard deviation is $sd(C_l) = 0.112$. As for the $C_d$, its free dynamics exhibits $\bar{C_d} = 3.46$ and $sd(C_d) = 0.0658$.
\input{FIG_V2/Fig8}

Figure \ref{FIG:Fig9} shows the controlled case in ideal measurement conditions (i.e. no noise), with the control parameters automatically selected using the BO algorithm. The simulations presented from now on include an initial unforced phase before activating the control at time instant $T_c = 50 \,c.u.$. Figure \ref{FIG:Fig9}($a,c,e$) shows the time histories of the controlled aerodynamic coefficients and the exogenous input provided to the system according to the MPC algorithm. \ref{FIG:Fig9}($b,d,f$) instead, displays the streamwise velocity component $u$ averaged over time windows of one shedding period of the fluidic pinball, in correspondence of the unforced, transient and steady control phases, respectively.

\input{FIG_V2/Fig9}

The control automatically selects an implementation law with the two rear cylinders in counter-rotation. The upper cylinder rotates clockwise ($b^2 = -b^3 <0$) with nearly constant rotational speed equal to the maximum values set in the optimization problem. This actuation mechanism corresponds to boat tailing, which has been previously observed and studied in various works on control of the wake of a fluidic pinball \citep{Raibaudo2020MLpinball,Li2022ExpGradientpinball}. 

This mechanism primarily aims to reduce the pressure drag of the fluidic pinball. Under unactuated conditions, as observed in plot $A$ of figure \ref{FIG:Fig9}, an extended recirculation region with low pressure and velocity forms behind the rear cylinders. The boat tailing redirects the shear layers of the top and bottom cylinder toward the pinball axis. This streamlining process of the wake increases the pressure behind the cylinders, delays the separation and energises the wake. Furthermore, the gap jet between the cylinders is substantially weakened. A detailed description of this mechanism can be found in \citet{Geropp2000BoatTailing}. A full wake stabilization is not achieved, i.e. the vortex shedding is not suppressed, due to the imposed constraints of $\abs{b^i}\leq1$. This is in line with the study of \cite{Maceda2021MLgradientpinball}, where wake stabilization in boat tailing is achieved with rotation of the rear cylinder of 2.375 or larger.

A condition of global optimum in terms of minimizing the drag coefficient would be achieved with an actuation that involves the rear cylinders in boat tailing and the front cylinder in a constant non-zero rotation, as observed in simulations \citep{Li2022ExpGradientpinball} and also experiments \citep{Raibaudo2020MLpinball}. The rotation of the front cylinder helps to reduce the low-velocity areas in the region behind the cylinders of the fluidic pinball, allowing for a slight reduction in drag. However, this condition is not achieved in the present case since the control target includes a condition of null $C_l$, with the consequence that any asymmetric flow actuation is penalised in the optimization process. Model predictive control, on the other hand, focuses its strategy towards reducing $C_l$ oscillations. \cite{Maceda2021MLgradientpinball} document oscillating lift coefficient with amplitude increasing with increasing $b^2,-b^3$ in the range $0-2$ if solely the rear cylinders are put in rotation. The self-tuned MPC converges to an actuation strategy to reduce $C_l$ oscillations based on phasor control with the front cylinder.

An examination of the behaviour of the drag coefficient ($C_d$) resulting from the implementation of the MPC control shows a decrease in its mean value, reducing it by $43.3\%$ compared with the unforced case. Moreover, the standard deviation of the drag coefficient is reduced as well, by $85.8\%$ during forced conditions compared with the unforced ones.

Figure \ref{FIG:Fig10} shows the out-of-plane vorticity in three different phases: uncontrolled, transient and post-transient the application of the control to the fluidic pinball wake. The MPC actuation is able to reduce the interaction between the upper and lower shear layers. In addition, the wake behind the fluidic pinball exhibits a more regular pattern in the post-transient phase. The wake meandering is strongly reduced, possibly due to the front stagnation point control with the front cylinder. The wake streamlining due to the boat tailing effect might also be contributing in this direction. This is also observed in the oscillating behaviour of the $C_l$, which becomes more regular and with less intense peaks once the control is activated. Indeed, in the controlled phase, the PSD of the $C_l$ presents a single very narrow peak concentrated at a Strouhal number of $\mathit{St}_D \approx 0.14$, resulting in a shedding period of $7.16 \,c.u.$, very similar to the predominant shedding frequency in the free response.
\input{FIG_V2/Fig10}

An analysis of the total power trend is required to assess if the MPC is able to achieve a net energy saving. The total power ($P_{tot}$) is characterised by the sum of two contributions: the first is directly related to the drag of the fluidic pinball ($P_d$), the second one is associated with the power required for actuation ($P_a$). 
Thus, the total power can be defined as
\begin{equation}
   P_{tot} = P_d + P_a = F_d U_\infty  + \sum_{i = 1}^{3} \frac{\tau^i b^i}{R}
\end{equation}
where $\tau^i$ is the torque acting on the $i^{th}$ cylinder of the fluidic pinball. The time history of power during the free and forced phases is given in figure \ref{FIG:Fig11}. It can be observed that under undisturbed conditions, the total power remains around an average value of $1.75$, with oscillations related only to the drag of the fluidic pinball. When moving to the controlled solution, the total power undergoes a sharp decrease in the transient phase thanks to the boat tailing. The total power stabilises at about $1.12$, experiencing a reduction of $36.31\%$ compared with the unforced phase. It is also observed that the average actuation power corresponds only to $12.13\%$ of the total power in the controlled phase. The average value of the actuation power is mainly related to the rear cylinders in boat tailing, while its oscillations are due to the front cylinder aiming to control the shedding, in order to reduce the lift fluctuations. It must be remarked though that this does not take into account possible inefficiencies that would arise in an implementation in a real environment.

\input{FIG_V2/Fig11}

A comparison between the proper orthogonal decomposition (POD) of the flow fields with and without control according to the MPC algorithm is also shown in figure \ref{FIG:Fig12}. 
It is specified that the POD has been performed on a dataset including both the transient and the stationary part of the control. On the left side of the figure, the spatial distributions of the first, third, and fifth out-of-plane vorticity modes are shown for both free and controlled cases. The plots on the right side of the figure show the eigenvalues of the first $100$ temporal modes. The eigenvalues are normalised with respect to their sum for the case without actuation. This allows a direct comparison of the mode energy in unforced and forced cases.

\input{FIG_V2/Fig12}

By observing the graphs related to the POD of the wake dataset under controlled conditions, it is noted that shedding inception is shifted upstream if compared with the unforced condition and is also more regular in the presence of boat tailing of the rear cylinders. Furthermore, the modes of the controlled case, compared with the free case, are characterised by lower energy, with the exception of the first two modes. This is ascribed to a partial wake stabilization, with reduced meandering. Mode $1$ spotlights indeed more defined vortical structures, with the crosswise width practically constant throughout the wake development. The fluidic boat tailing of the wake has indeed the effect of re-energizing the outer shear layers, redirecting them towards the pinball axis. The wake dynamics shift towards a less chaotic behaviour, as also observed in the more regular oscillations of the $C_l$ after activation of the control (figure \ref{FIG:Fig9}). A less rich dynamics results in a more compact POD eigenspectrum for the controlled case, with higher energy in the first two modes. The reduced crosswise oscillations are observed also in the structure of higher-order modes. Modes $3 - 6$ for the uncontrolled case embed the energy of higher-order harmonics and model the crosswise wake oscillations. In addition to the fluidic boat tailing due to the rotation of the rear cylinders, the stagnation point control enforced by the front cylinder has the effect of weakening such oscillations. This results in lower energy for the corresponding POD modes in the controlled case, and a more compact structure of their vortical features.
\subsubsection{The effects of measurements noise in the sensors}

The self-tuned control strategy with online LPR of the sensor signal is assessed in the presence of noise. The control results in terms of $C_d$ and $C_l$ and actuation are given in figure \ref{FIG:Fig13}.
We include as a reference the case of MPC without hyperparameter tuning, i.e. the control parameters are manually selected and all set to unity (except for the prediction/control window set to $3 c.u.$). The outcomes when the parameter selection is automatically performed using BO is reported for different noise levels. Without hyperparameter tuning, the performance of the control is quite poor for both $C_d$ and $C_l$. In this case, the mean drag coefficient is $3.0598$, i.e. significantly larger than the value of $\approx1.97$ achieved after hyperparameter tuning. Furthermore, MPC converges to an asymmetric controlled state, with a negative average lift coefficient. Quite surprisingly, this is achieved even if $R_{b^{2}}=R_{b^{3}}=1$, i.e. forcing symmetry in the hyperparameters does not necessarily lead to symmetric actuation. After hyperparameter tuning, the penalty weight of actuation (and actuation variability) on the rear cylinder is decreased with respect to the front cylinder (see Table \ref{TAB:TuningResults}). This allows stronger actuation on the rear cylinders, thus fostering strong fluidic boat tailing, with consequent strong drag reduction, accompanied by a weak action of the front cylinder to improve wake stability.

\input{FIG_V2/Fig13}

The cases analysed above are reproduced also in figure \ref{FIG:Fig14} in terms of trajectories of $C_d(t)$, $C_l(t)$ and $C_l(t-\tau)$, where $\tau$ corresponds to a quarter of the shedding period $T_{sh}$ of the fluidic pinball wake. The plots offer a description of the attractor onto which the wake dynamics of the fluidic pinball evolves, including the transitional phase from free to forced conditions. When the parameter tuning is performed, during the control $C_d$ experiences a remarkable reduction, whereas $C_l$ evolves on a dynamics similar to the unforced case, although less chaotic. Nevertheless, it is worth noting that the control of $C_l$ is not perfectly attained, as the model uncertainties hinder a complete description of the shedding dynamics of the fluidic pinball. Without hyperparameter tuning, in the first part of the transient the system transitions initially towards a state with a lower average $C_d$ (around 2.5) and $C_l$ oscillations of slightly lower intensity. The relatively high penalty on the actuation of the rear cylinders with respect to the front one, however, forces the system to avoid strong boat tailing and starts leveraging asymmetric actuation based on the front cylinder in the attempt to stabilise the wake through stagnation point control. The system finally converges to a limit cycle with a higher average $C_d$ and with large oscillations of the $C_l$ around a negative average.

In terms of the $C_d$, the control remains nearly unaffected by variations in measurement noise in the sensors. Regardless of the noise intensity, the posterior cylinders consistently maintain their boat tailing configuration, exhibiting maximum rotational velocity limited by the actuation constraints. On the other hand, the control of the lift coefficient demonstrates higher sensitivity to measurement noise. This is attributed to the increased susceptibility of the predictive model to disturbances in the initial conditions of online predictions. As the measurement noise intensifies, the frontal cylinder displays larger oscillations in the selected actuation law prescribed by the MPC. 

Nevertheless, LPR provides advantages to the algorithm, resulting in only marginal degradation of control performance. This increases the robustness of the application in realistic control scenarios characterised by high sensor noise.

\input{FIG_V2/Fig14}

\subsubsection{Symmetry in rear cylinder parameters}
This part is intended to provide more details about the choice of symmetry in the control parameters. Specifically, it concerns the components of the weight matrices that penalise the input expenditures and input variability of the rear cylinders during control, already introduced in \S~\ref{subsec:BayesOpt}.

The results of the control simulations with and without imposed symmetry conditions are presented in figure \ref{FIG:Fig15}. A certain bias is observed in the control of $C_l$ in the case without imposing symmetry conditions. This is related to an asymmetry in the rotation of the rear cylinders. However, under imposed symmetry, the control of the drag coefficient improves slightly, as the tangential speeds of rotation of the rear cylinders are both at the actuation limits, creating a stronger boat tailing.
This lower effectiveness of the optimization in the case without imposed symmetry can be explained by the relatively low importance of the $C_l$ fluctuations in the functional of BO in equation \eqref{Eqn:BOcost}, as discussed in \S~\ref{SubSec:BOresults}. The implementation of a law that binds the parameters of the rear cylinders reduces the total number of parameters to be tuned, leading to a less computationally expensive offline phase of the control. 

\input{FIG_V2/Fig15}
\section{Discussion and conclusion}
\label{sec:DiscussionFutureResearch}

In this work a noise-robust MPC algorithm that does not require user intervention neither for the plant modelling nor for the hyperparameters selection is proposed. The model used in the MPC optimization is selected from input-output data of the system under control using nonlinear system identification. The control parameters are automatically selected using a black-box optimization based on Bayesian methods. Additionally, the robustness of the algorithm is enhanced by the non-parametric smoothing technique LPR, which acts on the output data from the control sensors to address the presence of measurement noise.

The proposed control algorithm is successfully applied to the control of the wake of the fluidic pinball for drag reduction in a chaotic regime (specifically, for $\Rey_D = 150$) using solely aerodynamic forces to guide the control strategy. The methodology achieves good success in controlling a non-linear, chaotic and high-dimensional system. Being based on the MPC technique, the algorithm easily allows for the inclusion of non-linear constraints in the control and is very promising for applications where the control target is not a simple set-point stabilization. 

The proposed technique has shown to be highly robust to sensor measurement uncertainty, performing excellently even in realistic control scenarios characterised by a high level of noise. It must be remarked that, although rather realistic for force measurements, the Gaussian noise model might not be adequate if other sensing techniques are used. This may have an impact on plant model identification and LPR performance. However, the most attractive feature is that the method requires minimal user interaction, as the control parameters are automatically tuned by the BO algorithm according to the control target, taking into account also different fidelity levels for the plant predictions. The automatic procedure identified the most rewarding directions to optimise parameters with the aid only of a few numerical experiments. To control $C_d$ and $C_l$ of the fluidic pinball around a zero set point the MPC algorithm used a combination of two control mechanisms that have been previously considered in fluidic pinball control works: the boat tailing of the rear cylinders for drag reduction and the stagnation point control of the front cylinder for shedding control. This corresponded to a stronger penalty on the $C_d$ and a low penalty on the actuation cost of the rear cylinders. Without the framework proposed in this work, this process would have required parametric studies or suboptimal analysis, with the inherent difficulty of choosing relative weights between heterogeneous quantities.

It is worth highlighting that the control algorithm is currently being tested on a control case with relatively simple dynamics \citep{Deng2020ROMpinball, Deng2022ClustHierNetpinball}. This streamlines the process of coordinate selection and nonlinear system identification using the SINDYc technique. On the one hand, the aerodynamic force coefficients in separated wake flows exhibit features that render them suitable for plant modelling. On the other hand, the dynamics of the flow, although chaotic, evolve clearly on a low-dimensional attractor, thus simplifying the library selection for the application of SINDYc. For more complex flows (higher Reynolds number or with less clear features for straightforward coordinate identification), we envision that the challenge of MPC application will reside mostly in the coordinate selection and in the plant identification. Fast-paced advancements in grey- and black-box modelling are paving the way to interesting research pathways in these directions. The approach proposed here, however, will still be applicable and will benefit from such advancements. The offline optimization of the hyperparameters of the cost function can be applied independently of the method for coordinate selection or plant identification. Furthermore, the outcome will automatically adapt to the fidelity of the model plant when compared with the uncertainty of the measurements used to feed it.
\backsection[Funding]{The authors acknowledge the support from the research project PREDATOR-CM-UC3M. This project has been funded by the call "Estímulo a la Investigación de Jóvenes Doctores/as" within the frame of the Convenio Plurianual CM-UC3M and the V PRICIT (V Regional Plan for Scientific Research and Technological Innovation).}

\backsection[Acknowledgements]{The authors thank Prof. B. Noack, from Harbin Institute of Technology, and Prof. M. Morzynski, from Poznan University of Technology, for providing the code for the fluidic pinball simulations, and Dr. G. Cornejo Maceda for the post-processing code for force estimation. The authors also thank three anonymous referees for numerous useful comments that significantly improved this article.}

\backsection[Declaration of interests]{ The authors report no conflict of interest.}

\backsection[Data availability statement]{The code used in this work is made available at: \href{https://github.com/Lmarra1/Self-tuning-model-predictive-control-for-wake-flows}{https://github.com/Lmarra1/Self-tuning-model-predictive-control-for-wake-flows}. The dataset is openly available in Zenodo, accessible through the following link: \href{https://doi.org/10.5281/zenodo.10530019}{https://doi.org/10.5281/zenodo.10530019}}


\backsection[Author ORCIDs]{\\
\textcolor{orcidlogocol}{\orcidlink{0000-0001-9422-2808} 0000-0001-9422-2808} - Luigi Marra;\\
\textcolor{orcidlogocol}{\orcidlink{0000-0001-8537-9280} 0000-0001-8537-9280} - Andrea Meilán-Vila;\\
\textcolor{orcidlogocol}{\orcidlink{0000-0001-9025-1505} 0000-0001-9025-1505} - Stefano Discetti.}

\backsection[Author contributions]{Authors may include details of the contributions made by each author to the manuscript'}


\end{document}

%% file: FIG_V2/Fig1.tex
\begin{figure}
    \centerline{\includegraphics{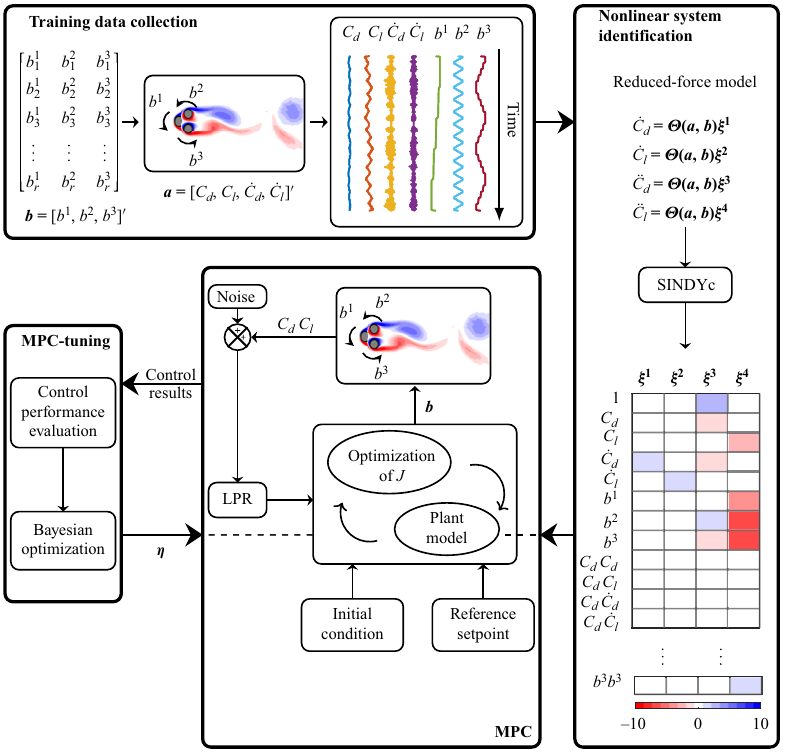}}
    \caption{\justifying The MPC algorithm schematic: dataset generation, creation of the predictive model and parameter tuning. The main block displays the closed-loop MPC scheme, including also LPR to mitigate the effects of sensor measurement noise.}
    \label{FIG:Fig1}
\end{figure}

%% file: FIG_V2/Fig2.tex
\begin{figure}
\centerline{\includegraphics{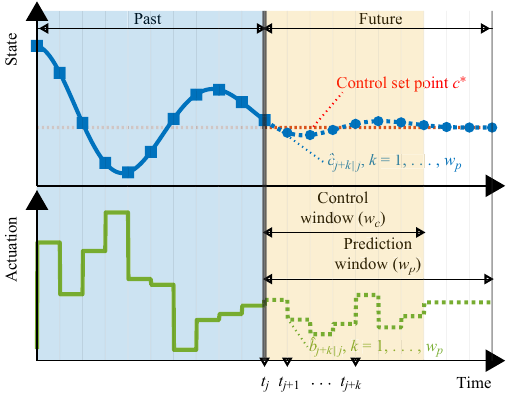}}
    \caption{\justifying Graphical representation of MPC strategy for stabilizing around a set point (horizontal dashed line). Past measurements (light-blue-shaded region) depict system state (blue lines with squares) and actuation (green line). The control window $w_c$ is shown in orange. Dashed lines indicate future state and actuation predictions. Blue circles represent a discrete sampling of the system state prediction. The continuous formulation allows non-mandatory discrete sampling and step-like actuation can be relaxed.}
    \label{FIG:Fig2}
\end{figure}

%% file: FIG_V2/Fig3.tex
\begin{figure}
    \centerline{\includegraphics{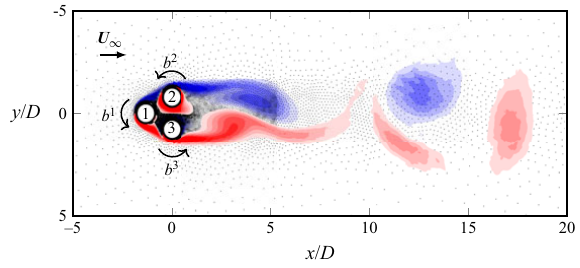}}
\caption{\justifying Domain of the incompressible two-dimensional DNS of the flow around the fluidic pinball. Front, top and bottom cylinders are labelled as $1$, $2$ and $3$, respectively. The rotational velocities of the cylinders are $b^1$, $b^2$ and $b^3$. The arrows indicate positive (counterclockwise) rotations. The background shows the $8633$ nodes grid used for the DNS. The contour colours indicate the out-of-plane vorticity.}
\label{FIG:Fig3}
\end{figure}

%% file: FIG_V2/Fig4.tex
\begin{figure}
    \centerline{\includegraphics{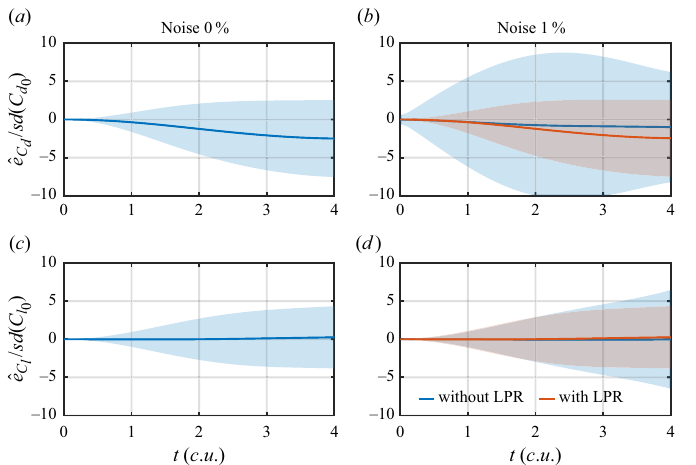}}
    \caption{\justifying Prediction performance of the force model obtained using SINDYc. The plots display the average value and the confidence region for $\sigma$ of the probability distribution of the normalized $C_d$ and $C_l$ prediction errors with respect to their standard deviation under unforced conditions. Panels ($a,c$) show the results in absence of measurement noise of the initial condition while ($b,d$) with $1\%$ of measurement noise. In the presence of noise, a comparison is shown with the results of predictions obtained using LPR estimation as an initial condition.
    }
    \label{FIG:Fig4}
\end{figure}

%% file: FIG_V2/Fig5.tex
\begin{figure}
    \centerline{\includegraphics{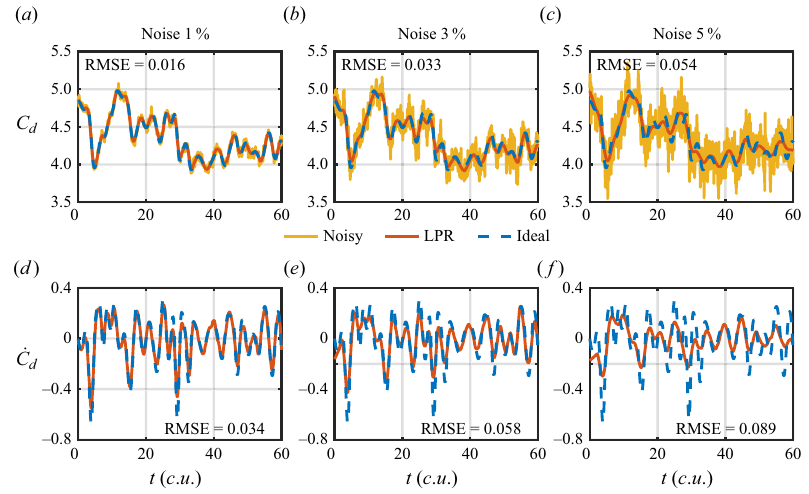}}
    \caption{\justifying Local polinomial regression applied to a $C_d$ time series within a time period of $60 c.u.$. This case corresponds to forced fluidic pinball dynamics. Panels ($a-c$) display $C_d$ estimation in the presence of increasing noise levels ($1\%, 3\%, 5\%$). Panels ($d-f$) show the LPR estimation of $\dot{C_d}$. Each plot includes the ideal and noisy time series, also including the RMSE of the LPR estimation.}
    \label{FIG:Fig5}
\end{figure}

%% file: FIG_V2/Fig6.tex
\begin{figure}
    \centerline{\includegraphics{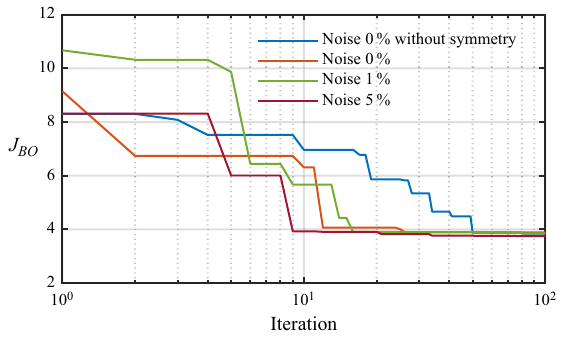}}
    \caption{\justifying Minimum observed $J_{BO}$ sampling history during MPC hyperparameter tuning. The $x$ axis of the plot is in logarithmic scale. The results of some control scenarios in the presence and absence of measurement noise (and symmetry in the parameters) are presented.}
    \label{FIG:Fig6}
\end{figure}

%% file: FIG_V2/Fig7.tex
\begin{figure}
\centerline{\includegraphics{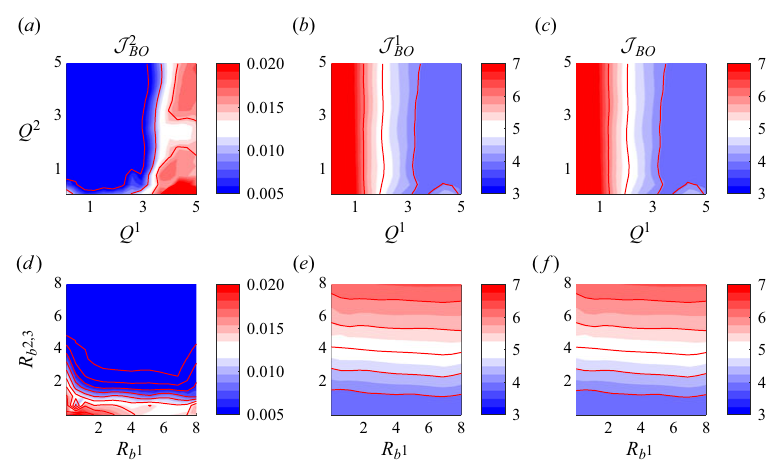}}
    \caption{\justifying  Representation of the cost functional $J_{BO}$ in equation (\ref{Eqn:BOcost}), the component due to $C_l$ control ($\mathcal{J}_{BO}^2$) and the component due to $C_d$ control ($\mathcal{J}_{BO}^1$). Contour plots are presented as functions of $Q^1$ and $Q^2$ ($a-c$) and $R_{b^1}$ and $R_{b^2} = R_{b^3} = R_{b^{2,3}}$ ($d-f$). The control case considers a $1\%$ measurement noise and symmetric weighting coefficients for the rear cylinders of the fluidic pinball. In the graphs, the remaining components of the hyperparameter vector $\boldsymbol{\eta}$ are fixed at their optimal values.}
    \label{FIG:Fig7}
\end{figure}

%% file: FIG_V2/Fig8.tex
\begin{figure}
  \centerline{\includegraphics{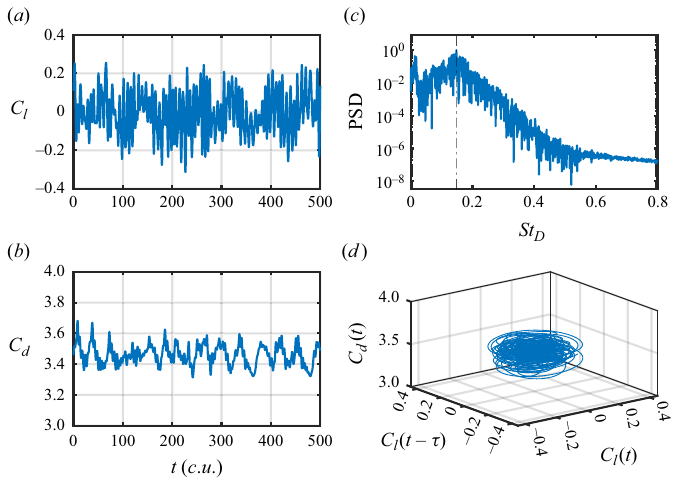}}
    \caption{\justifying Characteristics of the flow field around the fluidic pinball in undisturbed conditions. Panels ($a-b$) show the global lift and drag coefficients of the fluidic pinball. Plot ($c$) shows a PSD of the lift coefficient while ($d$) provides a representation of the trajectory in the time-delayed embedding space of the force coefficients $C_d(t)$, $C_l(t)$ and $C_l(t-\tau)$, where $\tau$ is a quarter of the shedding period of the fluidic pinball wake. The peak in the PSD is at a Strouhal number $\mathit{St}_D = 0.148$.}
    \label{FIG:Fig8}
\end{figure}

%% file: FIG_V2/Fig9.tex
\begin{figure}
    \hspace{0.2cm}
    \centerline{\includegraphics{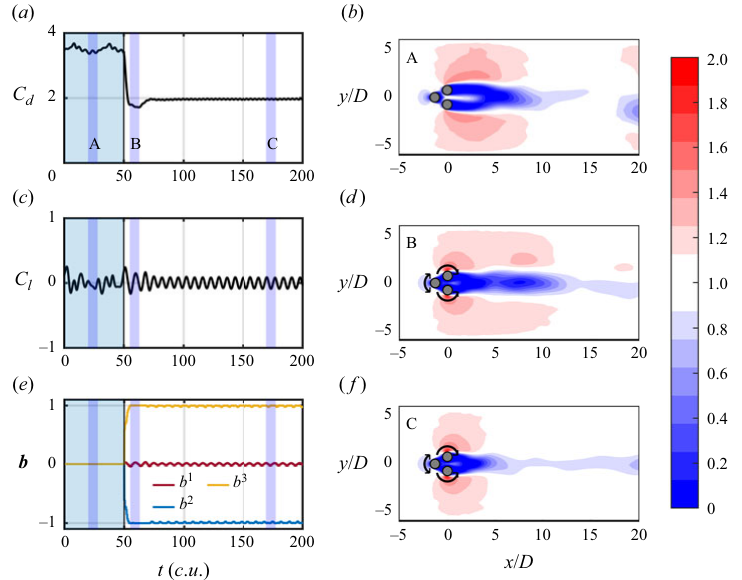}}
    \caption{Results of the control application in the absence of measurement noise in the sensors. Graphics ($a-c-e$) show the time histories of the $C_d$, $C_l$ and input vector $\boldsymbol{b}$ during an initial unforced phase and then during an active control phase. Graphics ($b-d-f$) show the mean value of the streamwise component of the velocity ($u$) in the time frames denoted A, B and C, respectively. Black arrows indicate the direction of rotation of the cylinders.}
    \label{FIG:Fig9}
\end{figure}

%% file: FIG_V2/Fig10.tex
\begin{figure}
    \centerline{
    \includegraphics{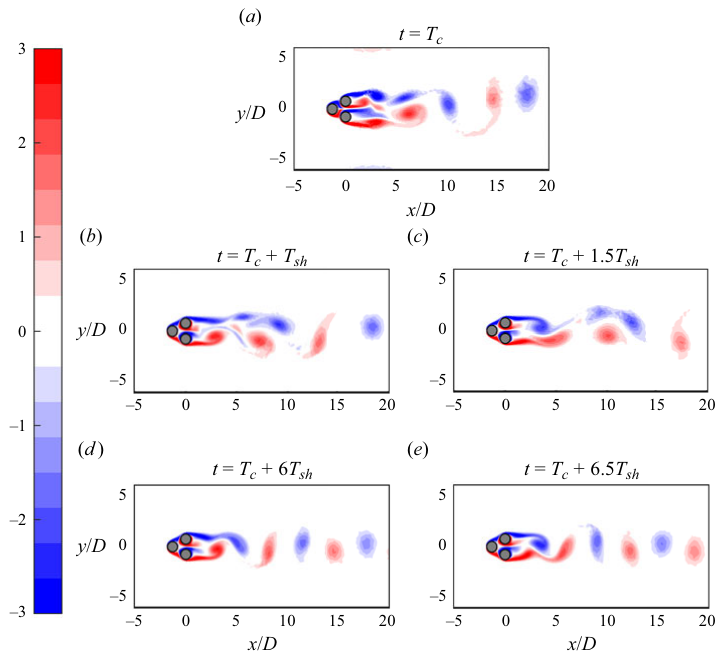}}
    \caption{\justifying Contour plot of the out-of-plane vorticity component of the flow around the fluidic pinball. Several instants corresponding to the phases of control onset, transient and post-transient phases are presented. Here $T_{sh}$ refers to the characteristic shedding period of the fluidic pinball in unforced conditions.}
    \label{FIG:Fig10}
\end{figure}

%% file: FIG_V2/Fig11.tex
\begin{figure}
 \centerline{\includegraphics{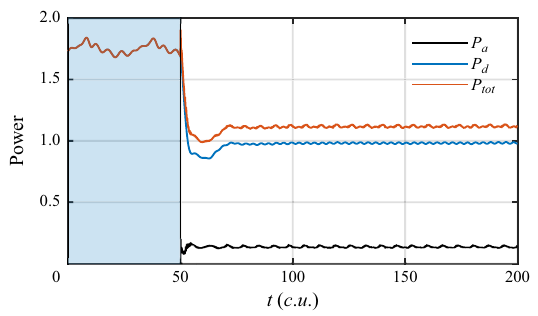}}
    \caption{Time histories of the power associated with drag, actuation and total power during the free and forced stages. Simulation of the fluidic pinball wake control in the absence of measurement noise.}
    \label{FIG:Fig11}
\end{figure}

%% file: FIG_V2/Fig12.tex
\begin{figure}
\centerline{\includegraphics{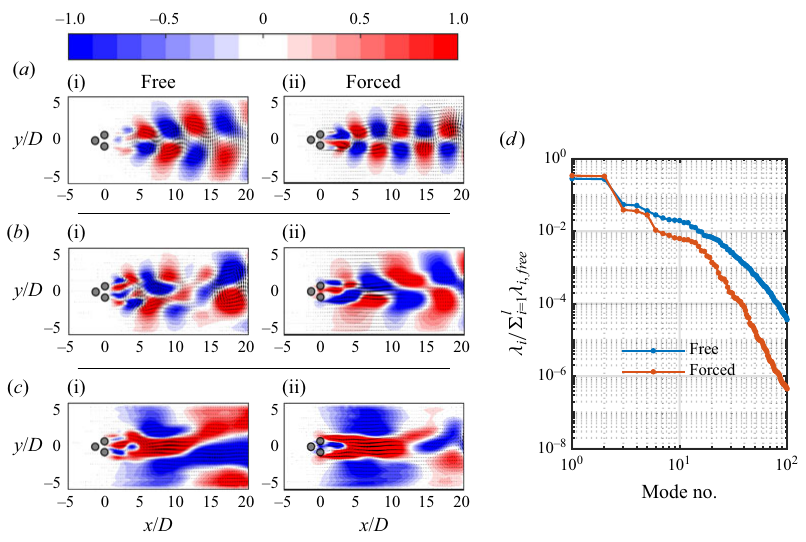}}
    \caption{\justifying Results of a POD applied to the flow field of the wake past fluidic pinball. Panels ($a-c$), from top to bottom depict the representation of out-of-plane vorticity for the spatial modes $1$, $3$ and $5$ under both free and forced conditions in accordance with the MPC architecture. Every plot also presents velocity vectors corresponding to each spatial mode. Plot ($d$) the squared singular values ($\lambda_i$) of the POD, normalized with respect to the sum of the squared singular values for the free case ($\lambda_{i,free}$). Proper orthogonal decomposition performed on a dataset consisting of $l = 1200$ snapshots of the fluidic pinball wake, including the transient in the forced case.}
        \label{FIG:Fig12}
\end{figure}

%% file: FIG_V2/Fig13.tex
\begin{figure}
    \centerline{\includegraphics{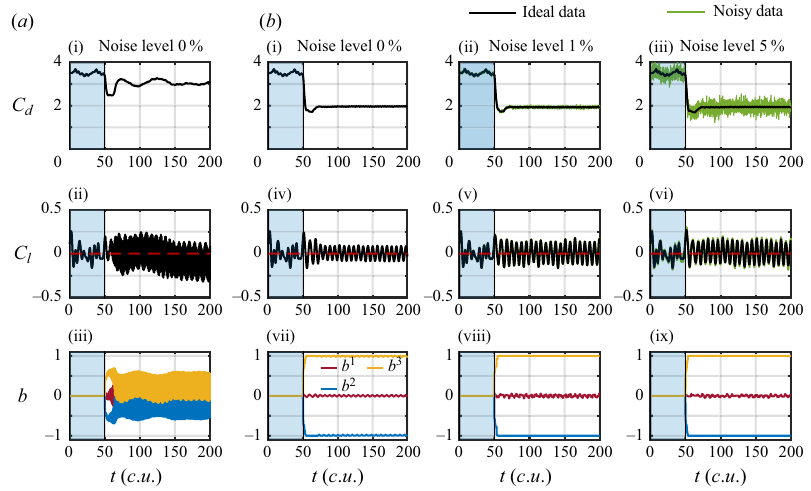}}
    \caption{\justifying Time series of $C_d$, $C_l$ and exogenous input $b^i$ (from top to bottom row, respectively) at free and forced conditions according to the MPC framework applied to the fluidic pinball. The panels in ($a$) show the results when the cost function parameters are hand selected and equal to the unity (with the exception of the prediction control window, set to $3  c.u.$). The panels in ($b$) show the results when BO is used for hyperparameter tuning. Local polynomial regression is included in the MPC framework in all cases where measurement noise is present.}
    \label{FIG:Fig13}
\end{figure}

%% file: FIG_V2/Fig14.tex
\begin{figure}
    \centerline{\includegraphics{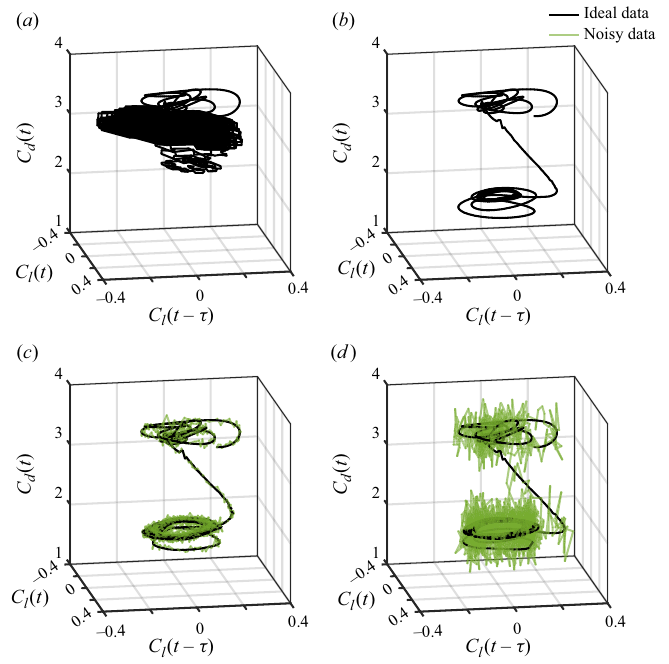}}
    \caption{ \justifying Trajectories in the time-delayed embedding space
of the force coefficients $C_d(t)$, $C_l(t)$ and $C_l(t-\tau)$, where $\tau$ is a quarter of the shedding period of the fluidic pinball wake. The panel in ($a$) shows the results when the cost function parameters are hand selected and equal to unity, while panels ($b-d$) show the result when BO is used for hyperparameter tuning. 
Cases ($a-b$) relate to ideal measurement conditions while cases ($c-d$) have noise of $1\%$ and $5\%$, respectively. Local polynomial is included in the MPC control framework in all cases where measurement noise is present.}
    \label{FIG:Fig14}
\end{figure}

%% file: FIG_V2/Fig15.tex
\begin{figure}
\centerline{
    \includegraphics{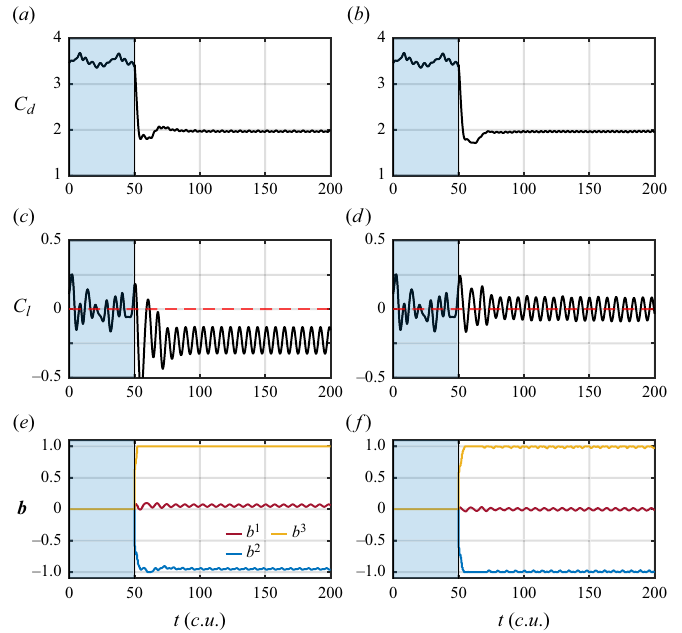}}
    \caption{Results of the application of MPC in terms of $C_d$, $C_l$ and input vector $\boldsymbol{b}$, without ($a-c-e$) and with ($b-d-f$) the imposition of symmetry in the parameters related to the rear cylinders. In both cases, all the parameters are tuned using the BO algorithm. Results of control in the absence of measurement noise in the sensors.}
\label{FIG:Fig15}
\end{figure}

%% file: Main.bbl
\begin{thebibliography}{75}
\expandafter\ifx\csname natexlab\endcsname\relax\def\natexlab#1{#1}\fi
\def\au#1{#1} \def\ed#1{#1} \def\yr#1{#1}\def\at#1{#1}\def\jt#1{\textit{#1}}
  \def\bt#1{#1}\def\bvol#1{\textbf{#1}} \def\vol#1{#1} \def\pg#1{#1}
  \def\publ#1{#1}\def\arxiv#1{#1}\def\org#1{#1}\def\st#1{\textit{#1}}

\bibitem[Ahuja {\em et~al.\/}(2007)Ahuja, Rowley, Kevrekidis, Wei, Colonius \&
  Tadmor]{Ahuja2007ControlLEvortices}
{\sc \au{Ahuja, S.}, \au{Rowley, C.}, \au{Kevrekidis, I.}, \au{Wei, M.},
  \au{Colonius, T.} \& \au{Tadmor, G.}} \yr{2007} Low-dimensional models for
  control of leading-edge vortices: equilibria and linearized models.  \bt{In
  {\em 45th AIAA Aerospace Sciences Meeting and Exhibit\/}},  \pg{p. 709}.

\bibitem[Allg\"ower {\em et~al.\/}(2004)Allg\"ower, Findeisen \&
  Nagy]{Allgower2004NMPCtheoryandapp}
{\sc \au{Allg\"ower, F.}, \au{Findeisen, R.} \& \au{Nagy, Z.~K.}} \yr{2004}
  \at{Nonlinear model predictive control: {F}rom theory to application}.
  \jt{J. Chin. Inst. Chem. Eng.}  \bvol{35}~(3),  \pg{299--316}.

\bibitem[Beintema {\em et~al.\/}(2020)Beintema, Corbetta, Biferale \&
  Toschi]{Beintema2020RLRBconvection}
{\sc \au{Beintema, G.}, \au{Corbetta, A.}, \au{Biferale, L.} \& \au{Toschi,
  F.}} \yr{2020}  \at{Controlling rayleigh–bénard convection via
  reinforcement learning}.  \jt{J. Turbul.}  \bvol{21}~(9-10),  \pg{585--605},
  \arxiv{arXiv: https://doi.org/10.1080/14685248.2020.1797059}.

\bibitem[Bewley {\em et~al.\/}(2001)Bewley, Moin \&
  Temam]{Bewley2001DNSpredictiveControlTurbulence}
{\sc \au{Bewley, T.~R.}, \au{Moin, P.} \& \au{Temam, R.}} \yr{2001}
  \at{Dns-based predictive control of turbulence: an optimal benchmark for
  feedback algorithms}.  \jt{J. Fluid Mech.}  \bvol{447},  \pg{179–225}.

\bibitem[Bieker {\em et~al.\/}(2020)Bieker, Peitz, Brunton, Kutz \&
  Dellnitz]{Bieker2019DeepMPC}
{\sc \au{Bieker, K.}, \au{Peitz, S.}, \au{Brunton, S.~L.}, \au{Kutz, J.~N.} \&
  \au{Dellnitz, M.}} \yr{2020}  \at{Deep model predictive control with online
  learning for complex physical systems}.  \jt{Theoretical and Computational
  Fluid Dynamics}  \bvol{34}~(10).

\bibitem[Blackburn \& Henderson(1999)]{Blackburn1999OscCyl}
{\sc \au{Blackburn, H.~M.} \& \au{Henderson, R.~D.}} \yr{1999}  \at{A study of
  two-dimensional flow past an oscillating cylinder}.  \jt{J. Fluid Mech.}
  \bvol{385},  \pg{255–286}.

\bibitem[B{\o}hn {\em et~al.\/}(2023)B{\o}hn, Gros, Moe \&
  Johansen]{Bohn2023LearningBasedMPC}
{\sc \au{B{\o}hn, E.}, \au{Gros, S.}, \au{Moe, S.} \& \au{Johansen, T.~A.}}
  \yr{2023}  \at{Optimization of the model predictive control meta-parameters
  through reinforcement learning}.  \jt{Eng. Appl. Artif. Intell.}  \bvol{123},
   \pg{106211}.

\bibitem[Brunton \& Noack(2015)]{Brunton2015ClosedLoopControl}
{\sc \au{Brunton, S.~L.} \& \au{Noack, B.~R.}} \yr{2015}  \at{Closed-loop
  turbulence control: Progress and challenges}.  \jt{Appl. Mech. Rev.}
  \bvol{67}~(5), 050801.

\bibitem[Brunton {\em et~al.\/}(2016{\natexlab{{\em a\/}}})Brunton, Proctor \&
  Kutz]{Brunton2016Sindy}
{\sc \au{Brunton, S.~L.}, \au{Proctor, J.~L.} \& \au{Kutz, J.~N.}}
  \yr{2016{\natexlab{{\em a\/}}}}  \at{Discovering governing equations from
  data by sparse identification of nonlinear dynamical systems}.  \jt{Proc.
  Natl. Acad. Sci. U.S.A.}  \bvol{113}~(15),  \pg{3932--3937}.

\bibitem[Brunton {\em et~al.\/}(2016{\natexlab{{\em b\/}}})Brunton, Proctor \&
  Kutz]{Brunton2016SINDYc}
{\sc \au{Brunton, S.~L.}, \au{Proctor, J.~L.} \& \au{Kutz, J.~N.}}
  \yr{2016{\natexlab{{\em b\/}}}}  \at{Sparse identification of nonlinear
  dynamics with control (sindyc)}.  \jt{IFAC-PapersOnLine}  \bvol{49}~(18),
  \pg{710--715}, 10th IFAC Symposium on Nonlinear Control Systems NOLCOS 2016.

\bibitem[Buche {\em et~al.\/}(2002)Buche, Stoll, Dornberger \&
  Koumoutsakos]{Buche2002MOEAcombustion}
{\sc \au{Buche, D.}, \au{Stoll, P.}, \au{Dornberger, R.} \& \au{Koumoutsakos,
  P.}} \yr{2002}  \at{Multiobjective evolutionary algorithm for the
  optimization of noisy combustion processes}.  \jt{IEEE Trans. Syst. Man
  Cybern.}  \bvol{32}~(4),  \pg{460--473}.

\bibitem[Camacho \& Alba(2013)]{Camacho2013MPC}
{\sc \au{Camacho, E.~F.} \& \au{Alba, C.~B.}} \yr{2013} {\em Model predictive
  control\/}.  \publ{Springer science \& business media}.

\bibitem[Castellanos {\em et~al.\/}(2022{\natexlab{{\em a\/}}})Castellanos,
  {Cornejo Maceda}, {de la Fuente}, Noack, Ianiro \&
  Discetti]{castellanos2022machine}
{\sc \au{Castellanos, R.}, \au{{Cornejo Maceda}, G.~Y.}, \au{{de la Fuente},
  I.}, \au{Noack, B.~R.}, \au{Ianiro, A.} \& \au{Discetti, S.}}
  \yr{2022{\natexlab{{\em a\/}}}}  \at{Machine-learning flow control with few
  sensor feedback and measurement noise}.  \jt{Phys. Fluids}  \bvol{34}~(4),
  \pg{047118}.

\bibitem[Castellanos {\em et~al.\/}(2022{\natexlab{{\em b\/}}})Castellanos,
  Michelis, Discetti, Ianiro \& Kotsonis]{castellanos2022reducing}
{\sc \au{Castellanos, R.}, \au{Michelis, T.}, \au{Discetti, S.}, \au{Ianiro,
  A.} \& \au{Kotsonis, M.}} \yr{2022{\natexlab{{\em b\/}}}}  \at{Reducing
  turbulent convective heat transfer with streamwise plasma vortex generators}.
   \jt{Exp. Therm. Fluid Sci.}  \bvol{134},  \pg{110596}.

\bibitem[Cetiner \& Rockwell(2001)]{Cetiner2001StreamwiseOscCyl}
{\sc \au{Cetiner, O.} \& \au{Rockwell, D.}} \yr{2001}  \at{Streamwise
  oscillations of a cylinder in a steady current. part 1. locked-on states of
  vortex formation and loading}.  \jt{J. Fluid Mech.}  \bvol{427},
  \pg{1–28}.

\bibitem[Chevalier {\em et~al.\/}(2007)Chevalier, H{\oe}pffner, {\AA}kervik \&
  Henningson]{Chevalier2007DelayTransition}
{\sc \au{Chevalier, M.}, \au{H{\oe}pffner, J.}, \au{{\AA}kervik, E.} \&
  \au{Henningson, D.~S.}} \yr{2007}  \at{Linear feedback control and estimation
  applied to instabilities in spatially developing boundary layers}.  \jt{J.
  Fluid Mech.}  \bvol{588},  \pg{163--187}.

\bibitem[Collis {\em et~al.\/}(2000)Collis, Chang, Kellogg \&
  Prabhu]{Collis2000LESandTurbControl}
{\sc \au{Collis, S.}, \au{Chang, Y.}, \au{Kellogg, S.} \& \au{Prabhu, R.}}
  \yr{2000} Large eddy simulation and turbulence control.  \bt{In {\em Fluids
  2000 Conference and Exhibit\/}},  \pg{p. 2564}.

\bibitem[{Cornejo Maceda} {\em et~al.\/}(2021){Cornejo Maceda}, Li, Lusseyran,
  Morzy{\'n}ski \& Noack]{Maceda2021MLgradientpinball}
{\sc \au{{Cornejo Maceda}, G.~Y.}, \au{Li, Y.}, \au{Lusseyran, F.},
  \au{Morzy{\'n}ski, M.} \& \au{Noack, B.~R.}} \yr{2021}  \at{Stabilization of
  the fluidic pinball with gradient-enriched machine learning control}.  \jt{J.
  Fluid Mech.}  \bvol{917},  \pg{A42}.

\bibitem[Corona \& {De Schutter}(2008)]{Corona2008CruiseContrMPC}
{\sc \au{Corona, D.} \& \au{{De Schutter}, B.}} \yr{2008}  \at{Adaptive cruise
  control for a smart car: A comparison benchmark for mpc-pwa control methods}.
   \jt{IEEE Trans. Control Syst. Technol.}  \bvol{16}~(2),  \pg{365--372}.

\bibitem[Cortelezzi \& Speyer(1998)]{Cortelezzi1998ControlTransition}
{\sc \au{Cortelezzi, L.} \& \au{Speyer, J.~L.}} \yr{1998}  \at{Robust
  reduced-order controller of laminar boundary layer transitions}.  \jt{Phys.
  Rev. E.}  \bvol{58}~(2),  \pg{1906}.

\bibitem[Deng {\em et~al.\/}(2020)Deng, Noack, Morzyński \&
  Pastur]{Deng2020ROMpinball}
{\sc \au{Deng, N.}, \au{Noack, B.~R.}, \au{Morzyński, M.} \& \au{Pastur,
  L.~R.}} \yr{2020}  \at{Low-order model for successive bifurcations of the
  fluidic pinball}.  \jt{J. Fluid Mech.}  \bvol{884},  \pg{A37}.

\bibitem[Deng {\em et~al.\/}(2022)Deng, Noack, Morzyński \&
  Pastur]{Deng2022ClustHierNetpinball}
{\sc \au{Deng, N.}, \au{Noack, B.~R.}, \au{Morzyński, M.} \& \au{Pastur,
  L.~R.}} \yr{2022}  \at{Cluster-based hierarchical network model of the
  fluidic pinball--cartographing transient and post-transient, multi-frequency,
  multi-attractor behaviour}.  \jt{J. Fluid Mech.}  \bvol{934},  \pg{A24}.

\bibitem[Duriez {\em et~al.\/}(2017)Duriez, Brunton \&
  Noack]{duriez2017machine}
{\sc \au{Duriez, T.}, \au{Brunton, S.~L.} \& \au{Noack, B.~R.}} \yr{2017} {\em
  Machine learning control-taming nonlinear dynamics and turbulence\/}, ,
  \vol{vol. 116}.  \publ{Springer}.

\bibitem[Edwards {\em et~al.\/}(2021)Edwards, Tang, Mamakoukas, Murphey \&
  Hauser]{Edwards2021LearningBasedMPC}
{\sc \au{Edwards, W.}, \au{Tang, G.}, \au{Mamakoukas, G.}, \au{Murphey, T.} \&
  \au{Hauser, K.}} \yr{2021} Automatic tuning for data-driven model predictive
  control.  \bt{In {\em 2021 IEEE International Conference on Robotics and
  Automation (ICRA)\/}},  \pg{pp. 7379--7385}. IEEE.

\bibitem[Fan {\em et~al.\/}(2020)Fan, Yang, Wang, Triantafyllou \&
  Karniadakis]{Fan2020reinforcement}
{\sc \au{Fan, D.}, \au{Yang, L.}, \au{Wang, Z.}, \au{Triantafyllou, M.~S.} \&
  \au{Karniadakis, G.~E.}} \yr{2020}  \at{Reinforcement learning for bluff body
  active flow control in experiments and simulations}.  \jt{Proc. Natl. Acad.
  Sci. U.S.A.}  \bvol{117}~(42),  \pg{26091--26098}.

\bibitem[Fan \& Gijbels(1996)]{fan1996LPRapplications}
{\sc \au{Fan, J.} \& \au{Gijbels, I.}} \yr{1996} {\em Local polynomial
  modelling and its applications\/}. {\em Monographs on statistics and applied
  probability series\/} 66.  \publ{London [u.a.]: Chapman \& Hall}.

\bibitem[Fr{\"o}hlich {\em et~al.\/}(2022)Fr{\"o}hlich, K{\"u}ttel, Arcari,
  Hewing, Zeilinger \& Carron]{Frohlich2022LearningBasedMPC}
{\sc \au{Fr{\"o}hlich, L.~P.}, \au{K{\"u}ttel, C.}, \au{Arcari, E.},
  \au{Hewing, L.}, \au{Zeilinger, M.~N.} \& \au{Carron, A.}} \yr{2022}
  Contextual tuning of model predictive control for autonomous racing.  \bt{In
  {\em 2022 IEEE/RSJ International Conference on Intelligent Robots and Systems
  (IROS)\/}},  \pg{pp. 10555--10562}. IEEE.

\bibitem[Gautier {\em et~al.\/}(2015)Gautier, Aider, Duriez, Noack, Segond \&
  Abel]{Gautier2015closed}
{\sc \au{Gautier, N.}, \au{Aider, J.-L.}, \au{Duriez, T.}, \au{Noack, B.~R.},
  \au{Segond, M.} \& \au{Abel, M.}} \yr{2015}  \at{Closed-loop separation
  control using machine-learning}.  \jt{J. Fluid Mech.}  \bvol{770},
  \pg{442–457}.

\bibitem[Gerhard {\em et~al.\/}(2003)Gerhard, Pastoor, K., Noack, Dillmann,
  Morzynski \& Tadmor]{Gerhard2003ModelBasedGalerkVortexShed}
{\sc \au{Gerhard, J.}, \au{Pastoor, M.}, \au{K., R.}, \au{Noack, B.},
  \au{Dillmann, A.}, \au{Morzynski, M.} \& \au{Tadmor, G.}} \yr{2003}
  Model-based control of vortex shedding using low-dimensional galerkin models.
   \bt{In {\em 33rd AIAA fluid dynamics conference and exhibit\/}},  \pg{p.
  4262}.

\bibitem[Geropp \& Odenthal(2000)]{Geropp2000BoatTailing}
{\sc \au{Geropp, D.} \& \au{Odenthal, H.-J.}} \yr{2000}  \at{Drag reduction of
  motor vehicles by active flow control using the coanda effect}.  \jt{Exp.
  Fluids}  \bvol{28}~(1),  \pg{74--85}.

\bibitem[Giordano \& Parrella(2008)]{Giordano2008NN4BandSelectionLLR}
{\sc \au{Giordano, F.} \& \au{Parrella, M.~L.}} \yr{2008}  \at{Neural networks
  for bandwidth selection in local linear regression of time series}.
  \jt{Comput. Stat. Data Anal.}  \bvol{52}~(5),  \pg{2435--2450}.

\bibitem[Green(2003)]{Green2003AviationEnvirChallenge}
{\sc \au{Green, J.~E.}} \yr{2003}  \at{Civil aviation and the environmental
  challenge}.  \jt{Aeronaut. J.}  \bvol{107}~(1072),  \pg{281–299}.

\bibitem[Gr{\"u}ne \& Pannek(2017)]{Grune2017NMPC}
{\sc \au{Gr{\"u}ne, L.} \& \au{Pannek, J.}} \yr{2017}  \at{Nonlinear model
  predictive control}.  \bt{In {\em Nonlinear model predictive control\/}}.
  \publ{Springer Cham}.

\bibitem[Gad-el Hak(2000)]{GadElHak2000FlowControl}
{\sc \au{Gad-el Hak, M.}} \yr{2000} {\em Flow Control: Passive, Active, and
  Reactive Flow Management\/}.  \publ{Cambridge University Press}.

\bibitem[Henson(1998)]{Henson1998NMCPstatus}
{\sc \au{Henson, M.~A.}} \yr{1998}  \at{Nonlinear model predictive control:
  current status and future directions}.  \jt{Comput. Chem. Eng.}
  \bvol{23}~(2),  \pg{187--202}.

\bibitem[Hewing {\em et~al.\/}(2020)Hewing, Wabersich, Menner \&
  Zeilinger]{Hewing2020LearningMPC}
{\sc \au{Hewing, L.}, \au{Wabersich, K.~P.}, \au{Menner, M.} \& \au{Zeilinger,
  M.~N.}} \yr{2020}  \at{Learning-based model predictive control: Toward safe
  learning in control}.  \jt{Annu. Rev. Control Robot. Auton. Syst.}
  \bvol{3}~(1),  \pg{269--296}.

\bibitem[Illingworth {\em et~al.\/}(2011)Illingworth, Morgans \&
  Rowley]{Illingworth2011ControlFlowReson}
{\sc \au{Illingworth, S.~J.}, \au{Morgans, A.~S.} \& \au{Rowley, C.~W.}}
  \yr{2011}  \at{Feedback control of flow resonances using balanced
  reduced-order models}.  \jt{J. Sound Vib.}  \bvol{330}~(8),  \pg{1567--1581}.

\bibitem[Kaiser {\em et~al.\/}(2018)Kaiser, Kutz \& Brunton]{Kaiser2018SINDyc}
{\sc \au{Kaiser, E.}, \au{Kutz, J.~N.} \& \au{Brunton, S.~L.}} \yr{2018}
  \at{Sparse identification of nonlinear dynamics for model predictive control
  in the low-data limit}.  \jt{Proc. R. Soc. Lond.}  \bvol{474}~(2219),
  \pg{20180335}.

\bibitem[Kaiser {\em et~al.\/}(2017)Kaiser, Noack, Spohn, Cattafesta \&
  Morzy{\'n}ski]{kaiser2017cluster}
{\sc \au{Kaiser, E.}, \au{Noack, B.~R.}, \au{Spohn, A.}, \au{Cattafesta, L.~N.}
  \& \au{Morzy{\'n}ski, M.}} \yr{2017}  \at{Cluster-based control of a
  separating flow over a smoothly contoured ramp}.  \jt{Theor. Comput. Fluid
  Dyn.}  \bvol{31},  \pg{579--593}.

\bibitem[Kim(2011)]{Kim2011ControlWallTurbulenceDragReduction}
{\sc \au{Kim, J.}} \yr{2011}  \at{Physics and control of wall turbulence for
  drag reduction}.  \jt{Philos. Trans. Royal Soc. A}  \bvol{369}~(1940),
  \pg{1396--1411}.

\bibitem[Kim \& Bewley(2007)]{Kim2007LinearFlowControl}
{\sc \au{Kim, J.} \& \au{Bewley, T.~R.}} \yr{2007}  \at{A linear systems
  approach to flow control}.  \jt{Annu. Rev. Fluid Mech.}  \bvol{39}~(1),
  \pg{383--417}.

\bibitem[Koumoutsakos {\em et~al.\/}(2001)Koumoutsakos, Freund \&
  Parekh]{Koumoutsakos2001GAJetMixing}
{\sc \au{Koumoutsakos, P.}, \au{Freund, J.} \& \au{Parekh, D.}} \yr{2001}
  \at{Evolution strategies for automatic optimization of jet mixing}.  \jt{AIAA
  J.}  \bvol{39}~(5),  \pg{967--969},  \arxiv{arXiv:
  https://doi.org/10.2514/2.1404}.

\bibitem[Lee(2011)]{Lee2011MPCreview}
{\sc \au{Lee, J.~H.}} \yr{2011}  \at{Model predictive control: Review of the
  three decades of development}.  \jt{Int. J. Control. Autom.}  \bvol{9}~(3),
  \pg{415--424}.

\bibitem[Lee {\em et~al.\/}(2001)Lee, Cortelezzi, Kim \&
  Speyer]{Lee2001ModelBasedTurbulentDragReduction}
{\sc \au{Lee, K.~H.}, \au{Cortelezzi, L.}, \au{Kim, J.} \& \au{Speyer, J.}}
  \yr{2001}  \at{Application of reduced-order controller to turbulent flows for
  drag reduction}.  \jt{Phys. Fluids}  \bvol{13}~(5),  \pg{1321--1330}.

\bibitem[Li {\em et~al.\/}(2022)Li, Cui, Jia, Li, Yang, Morzy{\'n}ski \&
  Noack]{Li2022ExpGradientpinball}
{\sc \au{Li, Y.}, \au{Cui, W.}, \au{Jia, Q.}, \au{Li, Q.}, \au{Yang, Z.},
  \au{Morzy{\'n}ski, M.} \& \au{Noack, B.~R.}} \yr{2022}  \at{Explorative
  gradient method for active drag reduction of the fluidic pinball and slanted
  ahmed body}.  \jt{J. Fluid Mech.}  \bvol{932},  \pg{A7}.

\bibitem[Little {\em et~al.\/}(2007)Little, Debiasi, Caraballo \&
  Samimy]{Little2007OpenLoopCavity}
{\sc \au{Little, J.}, \au{Debiasi, M.}, \au{Caraballo, E.} \& \au{Samimy, M.}}
  \yr{2007}  \at{Effects of open-loop and closed-loop control on subsonic
  cavity flows}.  \jt{Phys. Fluids}  \bvol{19}~(6), 065104.

\bibitem[Loiseau {\em et~al.\/}(2018)Loiseau, Noack \&
  Brunton]{Loiseau2018SparseredordModelling}
{\sc \au{Loiseau, J.~C.}, \au{Noack, B.~R.} \& \au{Brunton, S.~L.}} \yr{2018}
  \at{Sparse reduced-order modelling: sensor-based dynamics to full-state
  estimation}.  \jt{J. Fluid Mech.}  \bvol{844},  \pg{459–490}.

\bibitem[Monokrousos {\em et~al.\/}(2008)Monokrousos, Brandt, Schlatter \&
  Henningson]{Monokrousos2008TransitionDelay}
{\sc \au{Monokrousos, A.}, \au{Brandt, L.}, \au{Schlatter, P.} \&
  \au{Henningson, D.~S.}} \yr{2008}  \at{Dns and les of estimation and control
  of transition in boundary layers subject to free-stream turbulence}.
  \jt{Int. J. Heat Fluid Flow}  \bvol{29}~(3),  \pg{841--855}.

\bibitem[Morton {\em et~al.\/}(2018)Morton, Jameson, Kochenderfer \&
  Witherden]{Morton2018CylMPC}
{\sc \au{Morton, J.}, \au{Jameson, A.}, \au{Kochenderfer, M.~J.} \&
  \au{Witherden, F.}} \yr{2018} Deep dynamical modeling and control of unsteady
  fluid flows.  \bt{In {\em Advances in Neural Information Processing
  Systems\/} (ed. \ed{S.~Bengio, H.~Wallach, H.~Larochelle, K.~Grauman,
  N.~Cesa-Bianchi \& R.~Garnett})}, ,  \vol{vol.~31}.  \publ{Curran Associates,
  Inc.}

\bibitem[Nagarajan {\em et~al.\/}(2018)Nagarajan, Singha, Cordier \&
  Airiau]{Nagarajan2018OpenLoopCavity}
{\sc \au{Nagarajan, K.~K.}, \au{Singha, S.}, \au{Cordier, L.} \& \au{Airiau,
  C.}} \yr{2018}  \at{Open-loop control of cavity noise using proper orthogonal
  decomposition reduced-order model}.  \jt{Comput Fluids}  \bvol{160},
  \pg{1--13}.

\bibitem[Nair {\em et~al.\/}(2019)Nair, Yeh, Kaiser, Noack, Brunton \&
  Taira]{Nair2019ClusterFeedbControl}
{\sc \au{Nair, A.~G.}, \au{Yeh, Chi-An}, \au{Kaiser, E.}, \au{Noack, B.~R.},
  \au{Brunton, S.~L.} \& \au{Taira, K.}} \yr{2019}  \at{Cluster-based feedback
  control of turbulent post-stall separated flows}.  \jt{J. Fluid Mech.}
  \bvol{875},  \pg{345–375}.

\bibitem[Noack \& Morzyński(2017)]{Noack2017pinballToolkit}
{\sc \au{Noack, B.~R.} \& \au{Morzyński, M.}} \yr{2017}  \bt{The fluidic
  pinball—a toolkit for multiple-input multiple-output flow control (version
  1.0)}. {\em Tech. Rep.\/}.  \org{Tech. Rep. 02/2017. Chair of Virtual
  Engineering, Poznan University of Technology}.

\bibitem[Ouyang {\em et~al.\/}(2018)Ouyang, Zhou, Ma \& Tu]{Ouyang2018lprcontr}
{\sc \au{Ouyang, L.}, \au{Zhou, D.}, \au{Ma, Y.} \& \au{Tu, Y.}} \yr{2018}
  \at{Ensemble modeling based on 0–1 programming in micro-manufacturing
  process}.  \jt{Comput. Ind. Eng.}  \bvol{123},  \pg{242--253}.

\bibitem[Parkin {\em et~al.\/}(2014)Parkin, Thompson \&
  Sheridan]{Parkin2014OpenLoopBluffBody}
{\sc \au{Parkin, D.~J.}, \au{Thompson, M.~C.} \& \au{Sheridan, J.}} \yr{2014}
  \at{Numerical analysis of bluff body wakes under periodic open-loop control}.
   \jt{J. Fluid Mech.}  \bvol{739},  \pg{94–123}.

\bibitem[Pastur {\em et~al.\/}(2019)Pastur, Deng, Morzy{\'n}ski \&
  Noack]{Pastur2018ROMpinball}
{\sc \au{Pastur, L.~R.}, \au{Deng, N.}, \au{Morzy{\'n}ski, M.} \& \au{Noack,
  B.~R.}} \yr{2019} Reduced-order modeling of the fluidic pinball.  \bt{In {\em
  Chaotic Modeling and Simulation International Conference\/}},  \pg{pp.
  205--213}. Springer,  \publ{Cham: Springer International Publishing}.

\bibitem[Peitz {\em et~al.\/}(2020)Peitz, Otto \&
  Rowley]{Peitz2020DataDrivenKoopmanMPC}
{\sc \au{Peitz, S.}, \au{Otto, S.~E.} \& \au{Rowley, C.~W.}} \yr{2020}
  \at{Data-driven model predictive control using interpolated koopman
  generators}.  \jt{SIAM J. Appl. Dyn. Syst.}  \bvol{19}~(3),  \pg{2162--2193}.

\bibitem[Pinier {\em et~al.\/}(2007)Pinier, Ausseur, Glauser \&
  Higuchi]{Pinier2007ClosedLoopwing}
{\sc \au{Pinier, J.~T.}, \au{Ausseur, J.~M.}, \au{Glauser, M.~N.} \&
  \au{Higuchi, H.}} \yr{2007}  \at{Proportional closed-loop feedback control of
  flow separation}.  \jt{AIAA J.}  \bvol{45}~(1),  \pg{181--190}.

\bibitem[Poncet {\em et~al.\/}(2005)Poncet, Cottet \&
  Koumoutsakos]{Poncet2005ESwakes}
{\sc \au{Poncet, P.}, \au{Cottet, G.~H.} \& \au{Koumoutsakos, P.}} \yr{2005}
  \at{Control of three-dimensional wakes using evolution strategies}.  \jt{CR
  Mécanique}  \bvol{333}~(1),  \pg{65--77}, high-order methods for the
  numerical simulation of vortical and turbulent flows [special issue].

\bibitem[Qin \& Badgwell(2003)]{Qin2003SurveyMPC}
{\sc \au{Qin, S.~J.} \& \au{Badgwell, T.~A.}} \yr{2003}  \at{A survey of
  industrial model predictive control technology}.  \jt{Control Eng. Pract.}
  \bvol{11}~(7),  \pg{733--764}.

\bibitem[Rabault {\em et~al.\/}(2019)Rabault, Kuchta, Jensen, Réglade \&
  Cerardi]{rabault2019ANNandRLflowcontrol}
{\sc \au{Rabault, J.}, \au{Kuchta, M.}, \au{Jensen, A.}, \au{Réglade, U.} \&
  \au{Cerardi, N.}} \yr{2019}  \at{Artificial neural networks trained through
  deep reinforcement learning discover control strategies for active flow
  control}.  \jt{J. Fluid Mech.}  \bvol{865},  \pg{281–302}.

\bibitem[Raibaudo {\em et~al.\/}(2020)Raibaudo, Zhong, Noack \&
  Martinuzzi]{Raibaudo2020MLpinball}
{\sc \au{Raibaudo, C.}, \au{Zhong, P.}, \au{Noack, B.~R.} \& \au{Martinuzzi,
  R.~J.}} \yr{2020}  \at{Machine learning strategies applied to the control of
  a fluidic pinball}.  \jt{Phys. Fluids}  \bvol{32}~(1),  \pg{015108}.

\bibitem[Raković \& Levine(2018)]{RakovicV2018HandBookMPC}
{\sc \au{Raković, S.~V.} \& \au{Levine, W.~S.}} \yr{2018} {\em Handbook of
  Model Predictive Control\/}. {\em Control engineering\/} 1.  \publ{Cham:
  Springer International Publishing AG}.

\bibitem[Rowley \& Williams(2006)]{Rowley2006ControlOpenCavity}
{\sc \au{Rowley, C.~W.} \& \au{Williams, D.~R.}} \yr{2006}  \at{Dynamics and
  control of high-reynolds-number flow over open cavities}.  \jt{Annu. Rev.
  Fluid Mech.}  \bvol{38},  \pg{251--276}.

\bibitem[Sasaki \& Tsubakino(2020)]{Sasaki2020MPCcyl}
{\sc \au{Sasaki, Y.} \& \au{Tsubakino, D.}} \yr{2020}  \at{Designs of feedback
  controllers for fluid flows based on model predictive control and regression
  analysis}.  \jt{Energies}  \bvol{13}~(6),  \pg{1325}.

\bibitem[Schumm {\em et~al.\/}(1994)Schumm, Berger \&
  Monkewitz]{Schumm1994ControlWakeBluffBody}
{\sc \au{Schumm, M.}, \au{Berger, E.} \& \au{Monkewitz, P.~A.}} \yr{1994}
  \at{Self-excited oscillations in the wake of two-dimensional bluff bodies and
  their control}.  \jt{J. Fluid Mech.}  \bvol{271},  \pg{17–53}.

\bibitem[Shimomura {\em et~al.\/}(2020)Shimomura, Sekimoto, Oyama, Fujii \&
  Nishida]{Shimomura2020CLwing}
{\sc \au{Shimomura, S.}, \au{Sekimoto, S.}, \au{Oyama, A.}, \au{Fujii, K.} \&
  \au{Nishida, H.}} \yr{2020}  \at{Closed-loop flow separation control using
  the deep q network over airfoil}.  \jt{AIAA J.}  \bvol{58}~(10),
  \pg{4260--4270}.

\bibitem[Sipp(2012)]{Sipp2012OpenLoopCavity}
{\sc \au{Sipp, D.}} \yr{2012}  \at{Open-loop control of cavity oscillations
  with harmonic forcings}.  \jt{J. Fluid Mech.}  \bvol{708},  \pg{439–468}.

\bibitem[Snoek {\em et~al.\/}(2012)Snoek, Larochelle \&
  Adams]{Snoek2012BayesOpt}
{\sc \au{Snoek, J.}, \au{Larochelle, H.} \& \au{Adams, R.}} \yr{2012} Practical
  bayesian optimization of machine learning algorithms.  \bt{In {\em Advances
  in Neural Information Processing Systems\/} (ed. \ed{F.~Pereira, C.~J.
  Burges, L.~Bottou \& K.~Q. Weinberger})}, ,  \vol{vol.~25}.  \publ{Curran
  Associates, Inc.}

\bibitem[Steffen {\em et~al.\/}(2010)Steffen, Oztop \&
  Ritter]{Steffen2010LPRclosedloopRobotics}
{\sc \au{Steffen, J.}, \au{Oztop, E.} \& \au{Ritter, H.}} \yr{2010} Structured
  unsupervised kernel regression for closed-loop motion control.  \bt{In {\em
  2010 IEEE/RSJ International Conference on Intelligent Robots and Systems\/}},
   \pg{pp. 75--80}.

\bibitem[Tadmor {\em et~al.\/}(2011)Tadmor, Lehmann, Noack, Cordier, Delville,
  Bonnet \& Morzy{\'n}ski]{Tadmor2011WakeControl}
{\sc \au{Tadmor, G.}, \au{Lehmann, O.}, \au{Noack, B.~R.}, \au{Cordier, L.},
  \au{Delville, J.}, \au{Bonnet, J.~P.} \& \au{Morzy{\'n}ski, M.}} \yr{2011}
  \at{Reduced-order models for closed-loop wake control}.  \jt{Philosophical
  Transactions of the Royal Society A: Mathematical, Physical and Engineering
  Sciences}  \bvol{369}~(1940),  \pg{1513--1524}.

\bibitem[Thiria {\em et~al.\/}(2006)Thiria, Goujon-Durand \&
  Wesfreid]{Thiria2006WakeCyl}
{\sc \au{Thiria, B.}, \au{Goujon-Durand, S.} \& \au{Wesfreid, J.~E.}} \yr{2006}
   \at{The wake of a cylinder performing rotary oscillations}.  \jt{J. Fluid
  Mech.}  \bvol{560},  \pg{123--147}.

\bibitem[Tol {\em et~al.\/}(2019)Tol, De~Visser \&
  Kotsonis]{Tol2019ModelTransitionDelay}
{\sc \au{Tol, H.~J.}, \au{De~Visser, C.~C.} \& \au{Kotsonis, M.}} \yr{2019}
  \at{Experimental model-based estimation and control of natural
  tollmien--schlichting waves}.  \jt{AIAA J.}  \bvol{57}~(6),  \pg{2344--2355}.

\bibitem[Wand \& Jones(1994)]{Wand1994KernelSmoothing}
{\sc \au{Wand, M.~P.} \& \au{Jones, M.~C.}} \yr{1994} {\em Kernel Smoothing\/}.
  {\em Chapman \& Hall/CRC Monographs on Statistics \& Applied Probability\/}
  60.  \publ{Boca Raton, FL, U.S.: Chapman \& Hall}.

\bibitem[Wang {\em et~al.\/}(2023)Wang, Deng, {Cornejo Maceda} \&
  Noack]{Wang2023cluster}
{\sc \au{Wang, X.}, \au{Deng, N.}, \au{{Cornejo Maceda}, G.~Y.} \& \au{Noack,
  B.~R.}} \yr{2023}  \at{Cluster-based control for net drag reduction of the
  fluidic pinball}.  \jt{Phys. Fluids}  \bvol{35}~(2), 023601.

\bibitem[Wu {\em et~al.\/}(2018)Wu, Fan, Zhou, Li \& Noack]{Wu2018MLJet}
{\sc \au{Wu, Z.}, \au{Fan, D.}, \au{Zhou, Y.}, \au{Li, R.} \& \au{Noack,
  B.~R.}} \yr{2018}  \at{Jet mixing optimization using machine learning
  control}.  \jt{Exp. Fluids}  \bvol{59},  \pg{1--17}.

\end{thebibliography}
